\documentclass[a4paper]{article}

\usepackage{amssymb}
\usepackage{graphicx}
\usepackage{geometry}
\usepackage[cm]{fullpage}

\begin{document}

\title{Single-Spin CCD}
\author{T.~A. Baart$^{1}$\footnote{Contributed equally}, M. Shafiei$^{1*}$, T. Fujita$^{1}$, C. Reichl$^{2}$,\\ W. Wegscheider$^{2}$, L.~M.~K. Vandersypen$^{1}$\footnote{email: l.m.k.vandersypen@tudelft.nl}}
\maketitle

\begin{enumerate}
 \item QuTech and Kavli Institute of Nanoscience, TU Delft, 2600 GA Delft, The Netherlands
 \item Solid State Physics Laboratory, ETH Z\"{u}rich, 8093 Z\"{u}rich, Switzerland
\end{enumerate}

\textbf{Spin-based electronics or spintronics relies on the ability to store, transport and manipulate electron spin polarization with great precision~\cite{Prinz1995,Wolf2001,Zutic2004, Spintronics_insight2012}. In its ultimate limit, information is stored in the spin state of a single electron, at which point also quantum information processing becomes a possibility~\cite{Hanson2008,Taylor2005}. Here we demonstrate the manipulation, transport and read-out of individual electron spins in a linear array of three semiconductor quantum dots. First, we demonstrate single-shot read-out of three spins with fidelities of 97\% on average, using an approach analogous to the operation of a charge-coupled-device (CCD)~\cite{Boyle1970}. Next, we perform site-selective control of the three spins thereby writing the content of each pixel of this ``Single-Spin CCD''. Finally, we show that shuttling an electron back and forth in the array hundreds of times, covering a cumulative distance of 80 $\mu$m, has negligible influence on its spin projection. Extrapolating these results to the case of much larger arrays, points at a diverse range of potential applications, from quantum information to imaging and sensing.}\\

Past experiments have shown \emph{macroscopic} spin transport over distances exceeding 100 $\mu$m in clean bulk 2D semiconductors~\cite{Kikkawa1999, Crooker2005}. Furthermore, controlled single electron transport through semiconductor nanostructures~\cite{Grabert1992a} is now routine, with applications ranging from current standards to sensors and digital electronics~\cite{Ono2005}. Also the controlled transfer of individual electrons between nanostructures separated by several micron has been realized~\cite{McNeil2011,Hermelin2011}. However, the key combination of single-electron transport and spin preservation over large distances remains to be demonstrated.\\

A promising platform for moving around individual spins in a controlled manner is provided by analogy to a CCD~\cite{Boyle1970}. In a CCD, pockets of electrical charge are passed on along a capacitor array that acts as a shift-register, similar in spirit to a bucket-brigade. The pockets of charge arrive sequentially at the end of the array, where they are detected via a charge amplifier. This simple concept has enabled CCD camera’s containing millions of pixels with applications from consumer electronics to astronomy~\cite{Howell2006}. The creation of an analogous device that can operate and detect single spins would have powerful and diverse applications as well. For instance, it would not only enable reading out the outcome of a large quantum simulation or computation performed in a 2D array of spins~\cite{Taylor2005,Barthelemy2013}, but also could be used for coherent imaging at the single photon level~\cite{Vrijen2001,Fujita2013} or magnetic field sensing with 200 nm spatial resolution, using each single spin as a local probe. We term such a device a Single-Spin CCD. Its operation requires the ability to shuttle spins one by one along a chain of sites without disturbing the spin states, and to record the state of the spin at the end of the chain.\\

We have created a prototype Single-Spin CCD using a linear triple quantum dot array. The array is formed electrostatically in a 2D electron gas (2DEG) 85 nm below the surface, see Fig.~1a. Gate electrodes fabricated on the surface of a GaAs/AlGaAs heterostructure (see Methods) are biased with appropriate voltages to selectively deplete regions of the 2DEG and define the linear array of three quantum dots. Each dot is initially occupied with one electron, and each of the three electrons constitutes a single spin-$\frac{1}{2}$ particle. The main function of gates $LS$ and $RS$ is to set the tunnel coupling with the left and right reservoir, respectively. $D_{1}$ and $D_{2}$ control the interdot tunnel coupling (tuned to 0.8 and 0.5 GHz respectively) and $P_{1}$, $P_{2}$ and $P_{3}$ are used to set the electron number in each dot. The sensing dot (SD) next to the quantum dot array is used for non-invasive charge sensing using radiofrequency (RF) reflectometry to monitor the number of electrons in each dot~\cite{Barthel2010}. An in-plane magnetic field $B_{ext}$ = 3.51 T is applied to split the spin-up ($\uparrow$) and spin-down ($\downarrow$) states of each electron by the Zeeman energy ($E_{Z} \approx 87$ $\mathrm{\mu}$eV) defining a qubit in each of the dots. The electron temperature is 75 mK.\\

The CCD is initialized by loading $\uparrow$-spins from the right reservoir to every dot as described by the pulse sequence depicted by the blue and red arrows in Figs.~1bc. Once the desired spin manipulation has been performed (see below) in the (1,1,1)-regime, the read-out sequence is started. This sequence follows the reverse path from the loading sequence, with the addition of three read-out positions that are denoted by green circles. As in a CCD, the three electrons are pushed sequentially to the read-out site at the end of the chain. First, the right dot is tuned to the position of green circle nr. 3 in Fig.~1c. At this position, an excited spin-$\downarrow$ electron is allowed to tunnel to the reservoir, whilst a ground state spin-$\uparrow$ electron will remain in the dot. The nearby sensing dot is used to record whether or not the electron tunneled out, revealing its spin state~\cite{Elzerman2004a}. Next, we adjust the gate voltages to shuttle the center electron to the right dot for read out, at the position of green circle nr. 2. Afterwards we shuttle the left spin through the center dot to the right, and complete the three-spin read-out at the position of green circle nr. 1 in Fig.~1b. Details on the pulse scheme can be found in the Supplementary Information. \\

We test the operation of the Single-Spin CCD by preparing various combinations of the eight three-spin populations $\uparrow \uparrow \uparrow$ through $\downarrow \downarrow \downarrow$. To do so, we implement site-selective manipulation of each of the three spins exploiting the small difference in $g$-factors between the dots (Fig.~2i). We chirp a microwave electric field from well below to well above one of the three spin resonance frequencies, which results in adiabatic inversion of the corresponding spin via electric dipole spin resonance (EDSR)~\cite{Shafiei2013}, here with 76\% spin-$\downarrow$ initialization efficiency on average (see Supplementary Information). This amounts to `writing' the qubits of the Single-Spin CCD. Starting from $\uparrow \uparrow \uparrow$, we create in this way the spin states $\downarrow \uparrow \uparrow$, $\uparrow \downarrow \uparrow$ or $\uparrow \uparrow \downarrow$ and subsequently vary the waiting time in the (1,1,1)-configuration. During that time, the populations evolve as the spins relax back to the ground state $\uparrow \uparrow \uparrow$.\\ 

Figs.~2a-h show the eight three-spin-probabilities as a function of the waiting time for the three initial state preparations. In addition, data are shown where random spins are injected in each of the three dots. We see that the data follow closely the expected behavior (see caption), shown as solid lines, indicating proper operation of the Single-Spin CCD. More specifically, within the bounds of our measurement fidelities and statistical errors, there is no evidence of one measurement influencing the next, nor of mixing between adjacent spins driven by the exchange interaction~\cite{Nowack2011}. Such mixing would be visible as an initial rise in the $\downarrow$-fraction for a spin when its neighbor was prepared as $\downarrow$ (for instance, the $\downarrow \uparrow \uparrow$ and $\uparrow \uparrow \downarrow$ populations would initially rise for the case where $\uparrow \downarrow \uparrow$ is prepared).\\ 

This is the first demonstration of reading out multiple spins through the same reservoir. Read-out fidelities are on average 98.2 ($\pm$0.5)\% for spin-up and 95.8 ($\pm$0.3)\% for spin-down (see Supplementary Information for details). The spin-down fidelities are mainly limited by the $T_{1}$-decay ($\sim$10 ms) of the spins waiting to be read out and the finite measurement bandwidth through which spin-$\downarrow$ tunnel events are sometimes missed. The spin-up fidelities are likely limited by thermal broadening and/or electrical noise.\\

Next we examine the effect of shuttling electrons between dots on their spin projection. We anticipate three mechanisms that could in principle cause the spin-projection to change: (i) charge exchange with the reservoirs, (ii) spin-orbit (SO) interaction and (iii) hyperfine interaction with the nuclear spins of the quantum dot host material. Exchange with the reservoirs (i) is suppressed by applying precisely tuned pulse sequences and keeping the tunnel barriers of the reservoirs sufficiently closed. For the present gate voltage settings, we estimate this error to be $<10^{-5}$ per hop along the array (see Supplementary Information). The SO-interaction (ii) could affect the spin state, in a deterministic way, as it propagates through the array~\cite{Danon2009,Schreiber2011}. The direction of movement with respect to the crystal axis determines the magnitude and direction of the SO-field. We expect the SO-interaction to be largest for motion along the $[1\bar{1}0]$ axis and minimal along the interdot axis $[110]$~\cite{Golovach2004,Scarlino2014}, see (Fig.~1a). Furthermore, the small SO-field still originating from movement along this interdot axis will be \emph{parallel} to the external field and therefore will not influence the spin projection. We thus expect the effect of SO to be very small. The hyperfine interaction (iii) can cause random flips arising from the instantaneous unknown difference in perpendicular (relative to $B_{ext}$) hyperfine field, $\delta B_{\perp} $, between neighbouring dots. This effect is suppressed by the large $B_{ext}$ and estimated to be $< 1 \cdot 10^{-6}$ per shuttle event assuming  $\delta B_{\perp} <$ 7 mT (see Supplementary Information), and could even be corrected for using real-time Hamiltonian estimation~\cite{Shulman2014a}. \\

To verify experimentally whether it is possible to shuttle electrons while preserving their spin projection, we simulate a very large array using the triple dot device. Using the charge states from Fig.~1b, we load one random electron in the (0,0,1)-state and shuttle it back and forth many times to (1,0,0) by passing through (0,1,0) as depicted schematically in Fig.~3b. Each run of going back and forth constitutes a total of four jumps from one dot to the other. The time between hops is kept much longer than the single-spin $T_{2}^{*}$, which prevents coherent error accumulation in phase space. The effect of moving back and forth is therefore the same as traversing four dots in the same direction. In Fig.~3a, we vary both the total time spent in the CCD, $t_{CCD}$, and the number of interdot hops, $n_{hops}$, and read out the spin at the end. Fitting the data for each $t_{CCD}$ to a linear curve gives an average change in the spin-down fraction ranging between $-1.7 \cdot 10^{-6}$ and $+8.3 \cdot 10^{-6}$ per hop (see Supplementary Information). We compare the measured spin state after shuttling many times, with the spin state measured after the electron has shuttled back and forth only once in exactly the same total amount of time. The latter effectively constitutes a weighted $T_{1}$ measurement over the three dots. The fact that the triangular symbols in Fig.~3a fall exactly on the weighted $T_{1}$ decay, indicates that there is no sign of spin flips other than through spontaneous relaxation, even after more than 500 hops. Taking an interdot distance of 160 nm, this corresponds to a total distance of about 80 $\mu$m.\\ 

An important question to address is the scalability of this approach. The data of Fig.~3a shows that in the current experiment, the bottleneck is not the shuttling of the spins itself, but rather relaxation in the time the spins have to wait before they can be read out. More specifically, the limiting factors for this experiment are the times it takes to read out ($\sim$130$\mu$s), empty a dot ($\sim$75$\mu$s) and the time allowed for shuttling to the next dot dot ($\sim$10$\mu$s). The black curves in Fig.~3c show the predicted spin-down fidelity as a function of the CCD-length, extrapolated based on the present numbers. \\ 

With a few technical improvements involving additional pulse lines, larger interdot tunnel couplings, and a somewhat lower magnetic field, tunneling and emptying can occur on a ns timescale, the read-out time can be halved and the $T_1$ can be doubled (see Supplementary Information). This would give fidelities as shown in red in Fig.~3c. They allow one to read out 50 spins with $>83 \%$ fidelity. Adding another SD on the other side of the array will double the possible CCD length.\\

The read-out time can be shortened more dramatically via the inclusion of Pauli spin blockade (PSB) in the read-out scheme. PSB allows one to distinguish whether two neighboring spins are parallel or anti-parallel~\cite{Ono2002,Nowack2007}. PSB read-out within 1 $\mu$s for a fidelity of 97\% has been demonstrated in GaAs dots~\cite{Shulman2012}. To implement this method we use the right spin as an ancillary spin and one additional dot to quickly initialize this spin in the $\uparrow$-state using a so-called hot-spot where spins relax on a sub-$\mu$s timescale~\cite{Srinivasa2013,Yang2013}. Then we compare spin 1 and 2 using PSB and from that determine the state of spin 2. After discarding spin 1, we initialize spin 2 to spin-up, again using a hot-spot, and then bring it close to the charge detector by shuttling it to the rightmost dot. Spin 3 can now occupy dot 2 and its spin-state can be determined using PSB etc. This leads to the blue curve in Fig.~3c predicting fidelities above 88 \% for arrays larger than 1000 spins. Moving to a different host material such as Si or SiGe, $T_{1}$-times can reach seconds~\cite{Yang2013,Simmons2011}, boosting fidelities further (green curve). For such large $T_{1}$-times, we estimate that the read-out fidelity of the last spin that is read out in an array of 1000 dots, will decrease by only 0.07\% compared to the read-out fidelity of the first spin.\\ 

The high fidelity with which the spin projection can be preserved upon shuttling between dots thus allows scaling the Single-Spin CCD concept to linear arrays of hundreds of dots. Two-dimensional arrays would be a natural next step. Besides innovative gate designs, this would either require the use of weak spin-orbit materials like silicon, or (predictable) spin-orbit induced rotations need to be taken into account along at least one direction. Finally, in materials with negligible hyperfine coupling, such as $^{28}$Si enriched substrates~\cite{Veldhorst2014}, not only the spin projection but also the spin phase is expected to be preserved during shuttling. Such coherent single-spin shuttles allow qubits to be moved in the course of a quantum computation, an essential tool for powerful quantum computing architectures~\cite{Kielpinski2002,Taylor2005}.\\

 
\newpage

\textbf{Acknowledgements} The authors acknowledge useful discussions with the members of the Delft spin qubit team, sample fabrication by F.R. Braakman, and experimental assistance from M. Ammerlaan, A. van der Enden, J. Haanstra, R. Roeleveld, R. Schouten, M. Tiggelman and R. Vermeulen. This work is supported by the Netherlands Organization of Scientific Research (NWO) Graduate Program, the Intelligence Advanced Research Projects Activity (IARPA) Multi-Qubit Coherent Operations (MQCO) Program and the Swiss National Science Foundation.\\

\textbf{Author information}
T.A.B, M.S. and T.F. performed the experiment and analyzed the data, C.R. and W.W. grew the heterostructure, T.A.B., M.S., T.F. and L.M.K.V. contributed to the interpretation of the data and commented on the manuscript, and T.A.B. and L.M.K.V. wrote the manuscript. \\

\textbf{Competing financial interests}
The authors declare no competing financial interests.

 
\newpage

	\textbf{Figure 1 Linear array of three quantum dots and Single-Spin CCD operation.} \textbf{a} SEM image of a sample nominally identical to the one used for the measurements. Dotted circles indicate quantum dots, squares indicate Fermi reservoirs in the 2DEG, which are connected to ohmic contacts. The RF reflectance of the SD is monitored in order to determine the occupancies of the triple dot. \textbf{b,c } Charge stability diagram of the triple dot for two different values of $M$ (\textbf{b} $M=$ -42 mV, \textbf{c} $M=$ -56 mV). $L$, $M$ and $R$ are linear combinations of the voltages applied to gates $P_{1}$, $P_{2}$ and $P_{3}$, allowing us to partially compensate for cross-capacitances (see Supplementary Information). The occupancy of each dot is denoted by $(n,m,p)$ corresponding to the number of electrons in the left, middle and right dot respectively. The fading of the middle dot charge transition lines can be explained in a similar way as in~\cite{Yang2014} (black dashed lines indicate their positions). The CCD is initialized by loading three electrons from the right reservoir following first the blue arrows in b, and then pulsing $M$ to continue loading along the red arrows in c (analogous to a shift-register). We load $\uparrow$-spins by ramping slowly through the charge transition lines with the right reservoir (slow compared to the tunnel rate with the reservoir). Read-out occurs at the position of the green circles using spin-selective tunneling (see inset of c).

\newpage

	\textbf{Figure 2 Writing and reading out the Single-Spin CCD.} \textbf{a-h}, Measured three-spin populations as a function of waiting time between state preparation and read-out for four different state preparations. Starting from $\uparrow \uparrow \uparrow$ ($>95$\% initialization efficiency), states $\downarrow \uparrow \uparrow$ and $\uparrow \uparrow \downarrow$ are prepared by EDSR in the (1,1,1)-regime. State $\uparrow \downarrow \uparrow$ is created by applying EDSR in the (1,1,0)-regime, and then loading the third $\uparrow$-spin. For the fourth state preparation, the gate voltages are not ramped but pulsed across the relevant charge transitions, so that electrons with random spin will occupy the three dots. The contrast is limited to 0.8 mostly by imperfect adiabatic inversion and not by the read-out fidelities. Each datapoint is an average of 999 measurements (error bars two s.d.). Solid lines are products of the calculated single-spin probabilities based on the individual $T_{1}$ and initial spin-down probability of each dot. The fact that they overlap with the corresponding three-spin probabilities demonstrates that no (unintended) correlations were introduced between the spins. \textbf{i} Resonance frequency for each dot $i$ as a function of the magnetic field. A fit of the form $f_{res}=\frac{g_{1}^{i} \mu_{B} B}{h}+\frac{g_{3}^{i} \mu_{B} B^{3}}{h}$~\cite{Shafiei_thesis}, with $\mu_{B}$ the Bohr magneton, $h$ Planck’s constant, gives $g_{1}^{1}=-0.430 \pm 0.001 $, $g_{1}^{2}=-0.434 \pm 0.003$ and $g_{1}^{3}=-0.434 \pm 0.002$. Despite similar $g_{1}$-factors in dot 2 and 3, time-variations of the local nuclear field still give rise to stable configurations where we can selectively address the two dots (see Supplementary Information, also for $g_{3}^{i}$).

\newpage

\textbf{Figure 3 Preservation of the spin-projection during shuttling.} \textbf{a} Circles and squares: measured spin-down probability after $n_{hops}$ interdot hopping events for three values of the total time $t_{CCD}$. Black dashed lines are linear fits to the data. Diamonds: measured spin-down probability after shuttling back and forth just \textit{once}, as a function of the total shuttling time, $t_{CCD}$. The decay  time constant represents a weighted $T_{1}$ over all three dots ($T_{1,\mathrm{weighted}} = 10.6 \pm 0.5 $ ms). Triangles depict the average value of the shuttling data (circles and squares) for each value of $t_{CCD}$ (error bars two s.d.). \textbf{b} Schematic representation of the tunneling of a spin back and forth inside the CCD array. Arrows depict the backwards trajectory from the right to the left dot. Reversing this pathway returns the spin to the right dot. There are two separate stages with the electron in the middle dot, to prevent charge exchange with the reservoirs (see Supplementary Information). \textbf{c} Estimated spin-down measurement fidelity as a function of the length of the Single-Spin CCD for the current settings, improved conditions as described in the main text and for the proposed PSB-scheme in GaAs and SiGe. Solid lines indicate the fidelity for the spin that is read out last, dashed lines the average fidelity for the whole array. The spin-up fidelity is independent of the CCD size.

\newpage

\begin{figure*}[h!]
	\centering
	\includegraphics[width=1.0\textwidth]{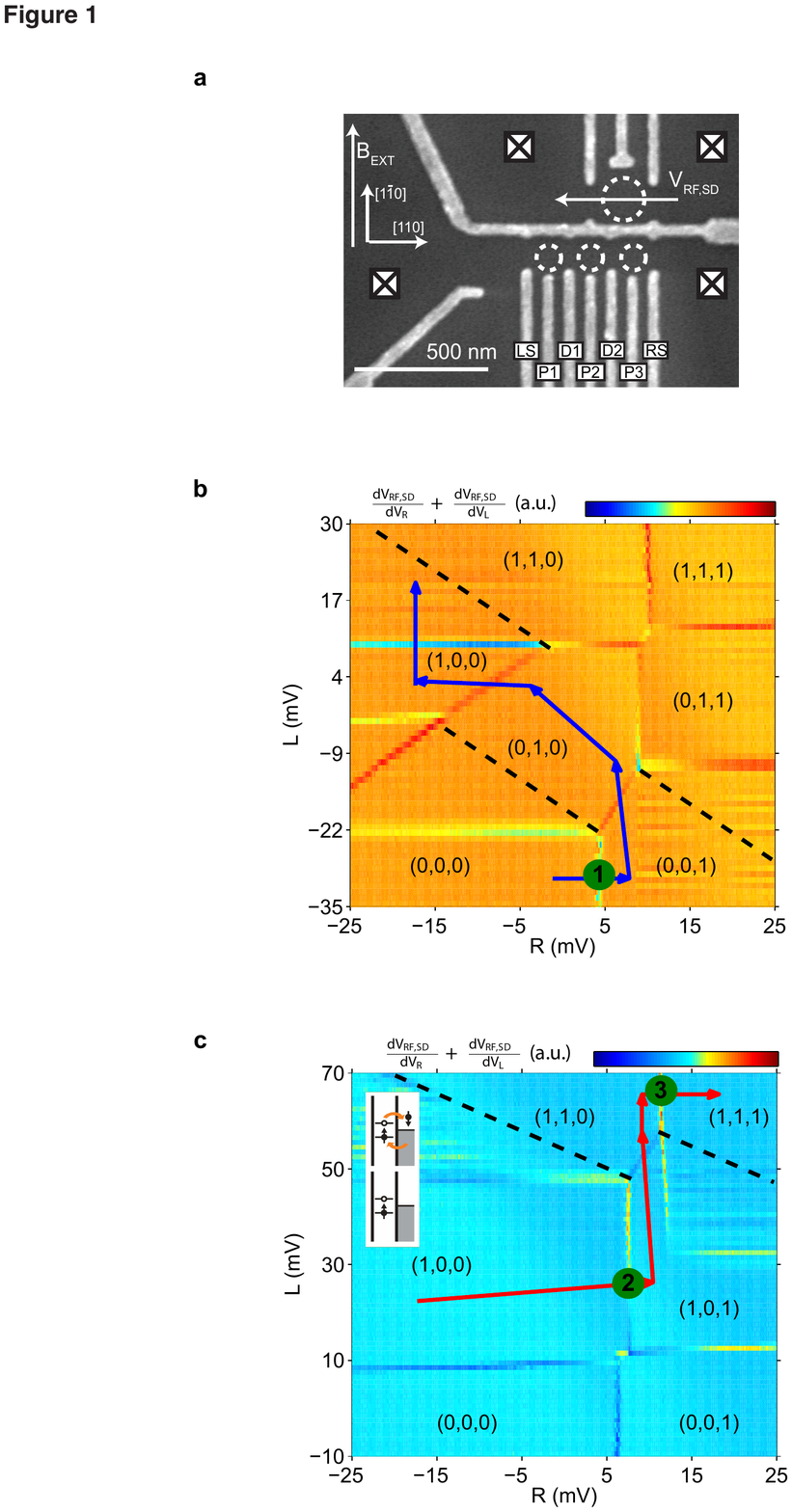}
\end{figure*}

\newpage

\begin{figure*}[h!]
	\centering
	\includegraphics[width=1.0\textwidth]{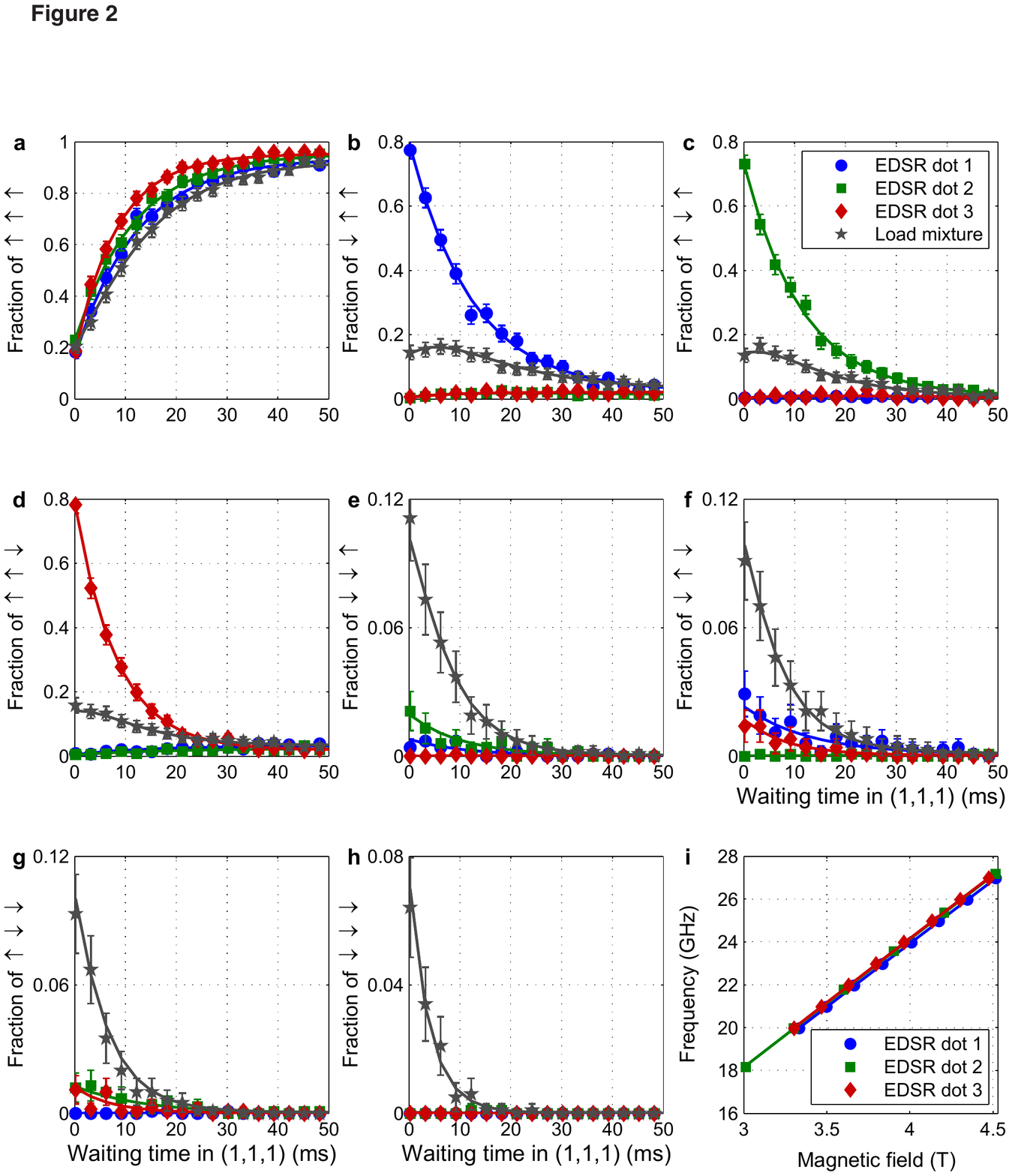}
\end{figure*}

\newpage

\begin{figure*}[h!]
	\centering
	\includegraphics[width=1.0\textwidth]{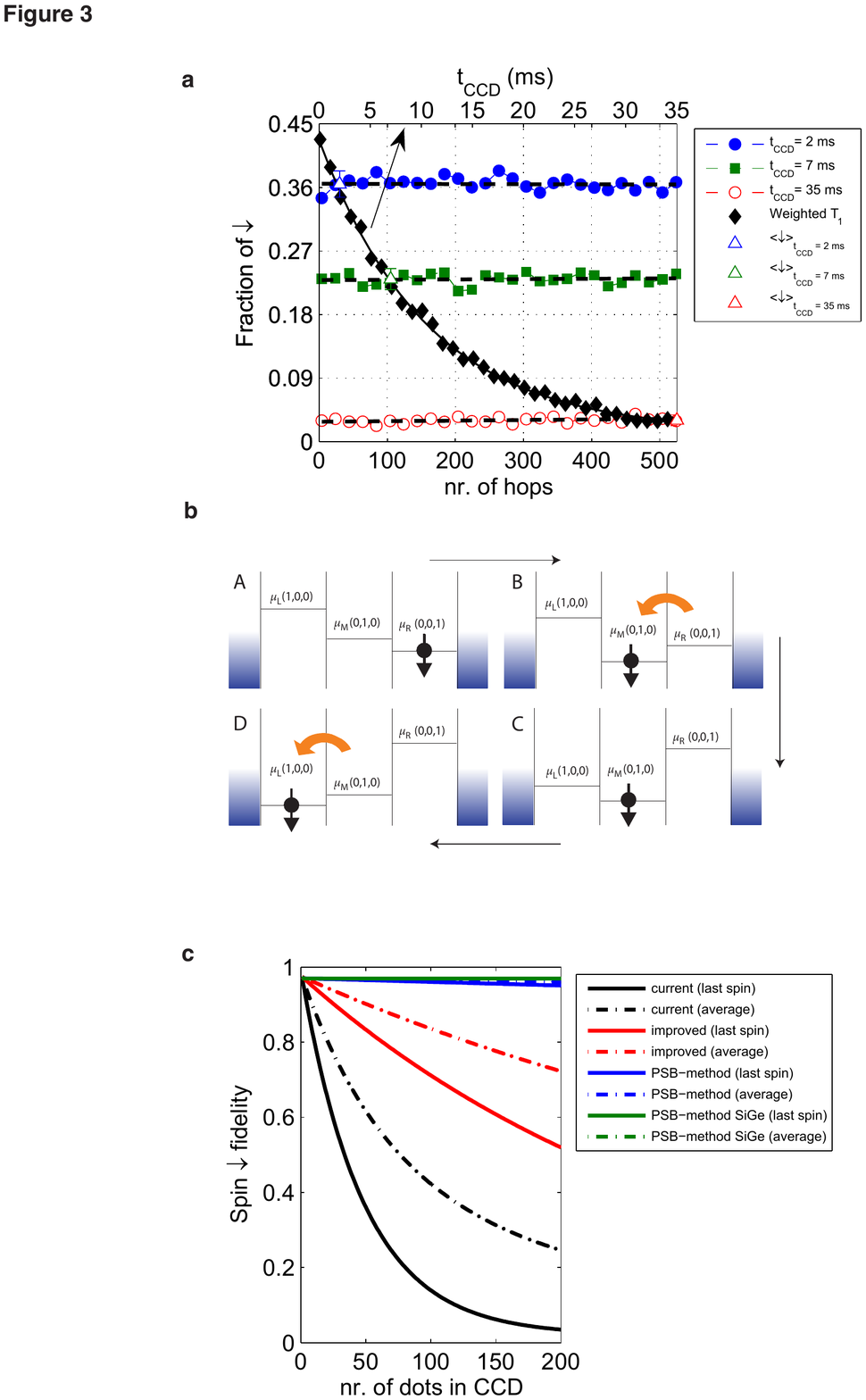}
\end{figure*}

\clearpage
\newpage

\renewcommand{\figurename}{Fig. S}
\renewcommand{\tablename}{Table S}

	\begin{centering}
		{\Large Supplementary Information for} \\ \vspace{0.2cm}
		{\Large \textbf{Single-Spin CCD}}\\
		\vspace{0.4cm}
		
		{\normalsize T.~A. Baart$^{1}$, M. Shafiei$^{1}$, T. Fujita$^{1}$,}\\  
		{\normalsize C. Reichl$^{2}$, W. Wegscheider$^{2}$, L.~M.~K. Vandersypen$^{1}$}\\
		\vspace{0.4cm}
		\normalsize{$^{1}$QuTech and Kavli Institute of Nanoscience, TU Delft, 2600 GA Delft, The Netherlands}\\
		\normalsize{$^{2}$Solid State Physics Laboratory, ETH Z\"{u}rich, 8093 Z\"{u}rich, Switzerland}\\
	\end{centering}
	
	\tableofcontents
	
	\newpage
	
	\section{Methods and materials}
	The experiment was performed on a $\mathrm{GaAs/Al_{0.25}Ga_{0.75}As}$ heterostructure grown by molecular-beam epitaxy, with a 85-nm-deep 2DEG with an electron density of $\mathrm{2.0 \cdot 10^{11}\ cm^{-2}}$ and mobility of $\mathrm{5.6 \cdot 10^{6}\ cm^{2} V^{-1} s^{-1}}$ at 4 K. The metallic (Ti–Au) surface gates were fabricated using electron-beam lithography. The device was cooled inside an Oxford Kelvinox 400HA dilution refrigerator to a base temperature of 45 mK. To reduce charge noise the sample was cooled while applying a positive voltage on all gates (ranging between 0 and 350 mV)~\cite{Long2006}. The device was tuned in the single-electron regime. The tunnel coupling at zero detuning between dot 1 and 2 was measured to be 0.8 GHz and 0.5 GHz between dot 2 and 3 using photon-assisted tunneling~\cite{Oosterkamp1998a}. Gates $P_{1}$, $P_{2}$, $P_{3}$ and $D_{2}$ were connected to homebuilt bias-tees ($RC$=470 ms), enabling application of d.c. voltage bias as well as high-frequency voltage excitation to these gates. RF reflectometry of the SD was performed using an LC circuit matching a carrier wave of frequency 110.35 MHz. The inductor is formed by a microfabricated NbTiN superconducting spiral inductor with an inductance of 3.0$\mu$H. The power of the carrier wave arriving at the sample was estimated to be -99 dBm. The carrier signal is only unblanked during read-out. The reflected signal was amplified using a cryogenic Weinreb CITLF2 amplifier and subsequently demodulated using homebuilt electronics. Real time data acquisition was performed using a FPGA DE0-Nano Terasic programmed to detect tunnel events using a Schmidt trigger. The microwaves were generated using a HP83650A connected to $P_{2}$ via a homemade bias-tee at room temperature. Voltage pulses to the gates were applied using a Tektronix AWG5014 (0-100\% risetime = 5 ns).
	
	\section{Calculation of the fidelities}
	In this part we discuss how we obtain the read-out fidelities from the data. We follow a similar approach as discussed in~\cite{Nowack2011} but extend the method to more dots, and incorporate some minor differences to comply with the CCD-scheme. We denote single spin read-out fidelity as $F_{i}^{j}$ where $j$ denotes the dot in order of the read-out (1 is the rightmost dot for the current sample, 2 is the center dot, and 3 the leftmost dot), and $i \in \{\uparrow,\downarrow\} $.
	
	\subsection{Analytic expressions for the fidelity}
	\label{sec:analytic_expression_for_the_fidelity}
	We start with the fidelity for the spin-up state, $F_{\uparrow}^{j}=1-\alpha^{j}$, where $\alpha^{j}$ is the probability that a step is detected in the SD signal although there was a spin-up electron in the quantum dot. This can occur if the SD signal exceeds the threshold even though the electron stayed in the dot (referred to as a signal processing error), or if the electron tunnels out of the dot due to thermal or electric field fluctuations. We can determine $\alpha^{j}$ directly for each quantum dot by initializing in the spin-up state and successively reading it out. In this case, the fraction of the cases where a spin-up electron is declared spin-down directly gives $\alpha^{j}$. This method assumes perfect initialization and thus gives an upper bound on $\alpha^{j}$. $\alpha^{j}$ might be slightly different for the different dots due to differences in the tunnel barrier with the right reservoir for each charge state, the length of the read-out stages, the thresholds and the signal heights. Except for these slight differences, $F_{\uparrow}^{j}$ is independent of the size of the array. There are two effects that may influence a spin-up state during the read-out sequence: (1) back-action from the SD or (2) thermal excitation. In the CCD scheme, each electron will experience a similar amount of back-action from the SD (an electron far away in an array will not yet experience back-action). Thermal excitation is negligible for the current energy scales: $E_{z} \approx 90~\mu$eV and $k_{B}T \approx 6.5~\mu$eV. In thermal equilibrium, an initially perfect spin-up electron will have a probability of $\sim 10^{-6}$ to be in the spin-down state (Boltzmann distribution). This is a negligible correction for our current spin-up fidelities.\\
	
	Next, we estimate the fidelity for the spin-down state based on the relaxation time $T_{1}^{j}$, the duration of the read-out stage, $T_{R}^{j}$, the electron tunneling rates (denoted by $\Gamma_{\sigma}^{i,j}$, where $\sigma \in \mathrm{\{in,out}\}$ represents tunneling in or out of the quantum dot) and an analysis of the measurement bandwidth.\\
	
	The spin-down fidelity for the first dot, $j=1$, is $F_{\downarrow}^{1} = \Xi^{1}+\Lambda^{1}$. $\Xi^{j}$ describes the probability that a spin-down electron tunnels out during the read-out stage before it relaxes. $\Lambda^{j}$ describes the probability that a spin-down electron relaxes during the read-out stage, but still tunnels out because of the nonzero $\Gamma_{out}^{\uparrow,j}$.\\
	
	First we assume infinite measurement bandwidth. In that case $\Xi^{j}$ can be defined as follows:
	\begin{equation}
	\Xi^{j} = \int\limits_{0}^{T_{R}^{j}} \Gamma_{out}^{\downarrow,j} P_{\downarrow}(t) dt
	\end{equation} 
	where $\Gamma_{out}^{\downarrow,j} P_{\downarrow}(t)$ is the probability density function for a spin-down electron to tunnel out at time $t$. The probability for the electron to be spin-down, $P_{\downarrow}(t)$, or spin-up, $P_{\uparrow}(t)$, at time $t$ follows from rate equations: (1) $\frac{dP_{\uparrow}}{dt} = -\Gamma_{out}^{\uparrow}P_{\uparrow} + \left( 1/T_{1} \right) P_{\downarrow} $ and (2) $\frac{dP_{\downarrow}}{dt} = -(\Gamma_{out}^{\downarrow} + 1/T_{1}) P_{\downarrow} $. These equations describe the probabilities \textit{before} the read-out event occurred (after which a spin-up electron will fill the dot with rate $\Gamma_{in}^{\uparrow,j}$). The initial conditions are $P_{\uparrow} (t=0) = 0$ and $P_{\downarrow} (t=0) = 1$. From this we find $P_{\uparrow} (t) = \frac{1}{1+T_{1}(\Gamma_{out}^{\downarrow}-\Gamma_{out}^{\uparrow})} (e^{-\Gamma_{out}^{\uparrow}t} - e^{-(\frac{1}{T_{1}}+\Gamma_{out}^{\downarrow})t} )$ and $ P_{\downarrow}(t) = e^{-(\frac{1}{T_{1}}+\Gamma_{out}^{\downarrow})t }$. \\
	
	Next we add the measurement bandwidth to the calculation. A tunnel event leads to a pulse in the SD signal. The duration of this pulse is determined by $\Gamma_{in}^{\uparrow,j}$. A pulse with duration $\tau$ will be detected with probability $B^{j} (\tau) $, which depends on the measurement bandwidth (see below). Therefore the probability that a pulse which is caused by an electron that tunnels out is detected is: 
	
	\begin{equation}
	\Xi^{j} = \int\limits_{0}^{T_{R}^{j}} \Gamma_{out}^{\downarrow,j} P_{\downarrow}(t) dt \int\limits_{0}^{\infty} \Gamma_{in}^{\uparrow,j} e^{-\Gamma_{in}^{\uparrow,j} \tau} B^{j}(\tau) d\tau 
	\label{eq:Xi}
	\end{equation} 
	
	We can extend the limit of the second integral to $\infty$ because we also count events where an electron did not yet tunnel back during the read-out time as spin-down. Eq. (\ref{eq:Xi}) assumes that we detect every electron leaving the dot with a 100\% probability. This assumption is not correct for tunnel events occurring very close to the end of the read-out stage. Based on the measured bandwidth (see below) we can assume that every electron leaving the dot at least 0.7 $\mu$s before the end  of the read-out stage will be detected. This gives the following lower bound for $\Xi^{j}$: 
	
	\begin{equation}
	\Xi^{j} = \int\limits_{0}^{T_{R}^{j}-0.7 \mu s} \Gamma_{out}^{\downarrow,j} P_{\downarrow}(t) dt \int\limits_{0}^{\infty} \Gamma_{in}^{\uparrow,j} e^{-\Gamma_{in}^{\uparrow,j} \tau} B^{j}(\tau) d\tau 
	\end{equation} 
	
	At the end of each read-out we always empty the right dot before going to the next read-out position so subsequent read-out stages always have the same charge configuration.\\
	
	The probability that a spin-down electron relaxes during the read-out stage before it could tunnel out but that nevertheless a step is detected due to the spin-up electron tunneling out is given by
	
	\begin{equation}
	\Lambda^{j} = \int\limits_{0}^{T_{R}^{j} -0.7 \mu s} \Gamma_{out}^{\uparrow,j} P_{\uparrow}(t) dt \int\limits_{0}^{\infty} \Gamma_{in}^{\uparrow,j} e^{-\Gamma_{in}^{\uparrow,j} \tau} B^{j}(\tau) d\tau 
	\end{equation} 
	
	where $\Gamma_{out}^{\uparrow,j} P_{\uparrow}(t)$ is the probability density function that the initial spin-down electron has relaxed to spin-up before tunneling out (recall we consider here the case that  $P_{\uparrow} (t=0) = 0$). \\
	
	The spin-down fidelity for dots with $j>1$ are additionally affected by relaxation during the previous read-out stages including the emptying of all dots $<j$, and the time it takes to shuttle from their position in the array to the read-out position, which adds a shuttling time. The total waiting time before the $j^{th}$ dot is read-out is denoted by $T_{wait}^{j}$.
	This relaxation occurs with probability $\eta^{j} = \int_{0}^{T_{wait}^{j}} \frac{1}{T_{1}^{j}} e^{-t/T_{1}^{j}} dt$. Therefore the spin down fidelity for the quantum dots with $j>1$ is
	\begin{equation}
	F_{\downarrow}^{j}=(\Xi^{j}+\Lambda^{j})(1-\eta^{j}) + \eta^{j}\alpha^{j}
	\end{equation} 
	The last term accounts for the case that the spin down electron of the $j^{th}$ dot relaxes to spin up during $T_{wait}^{j}$ but in the end is still detected as spin down. For the experiment the following values were used: $T_{R}^{1}=T_{R}^{2}$=130 $\mu$s, $T_{R}^{3}$=300 $\mu$s, $T_{wait}^{2}$ = 222.15 $\mu$s and $T_{wait}^{3}$ = 514.3 $\mu$s. These values were chosen to optimize the sum of the three spin-down fidelities, whilst keeping the spin-up fidelity high. \\
	
\begin{figure*}[h!]
		\centering
		\includegraphics[width=0.5\textwidth]{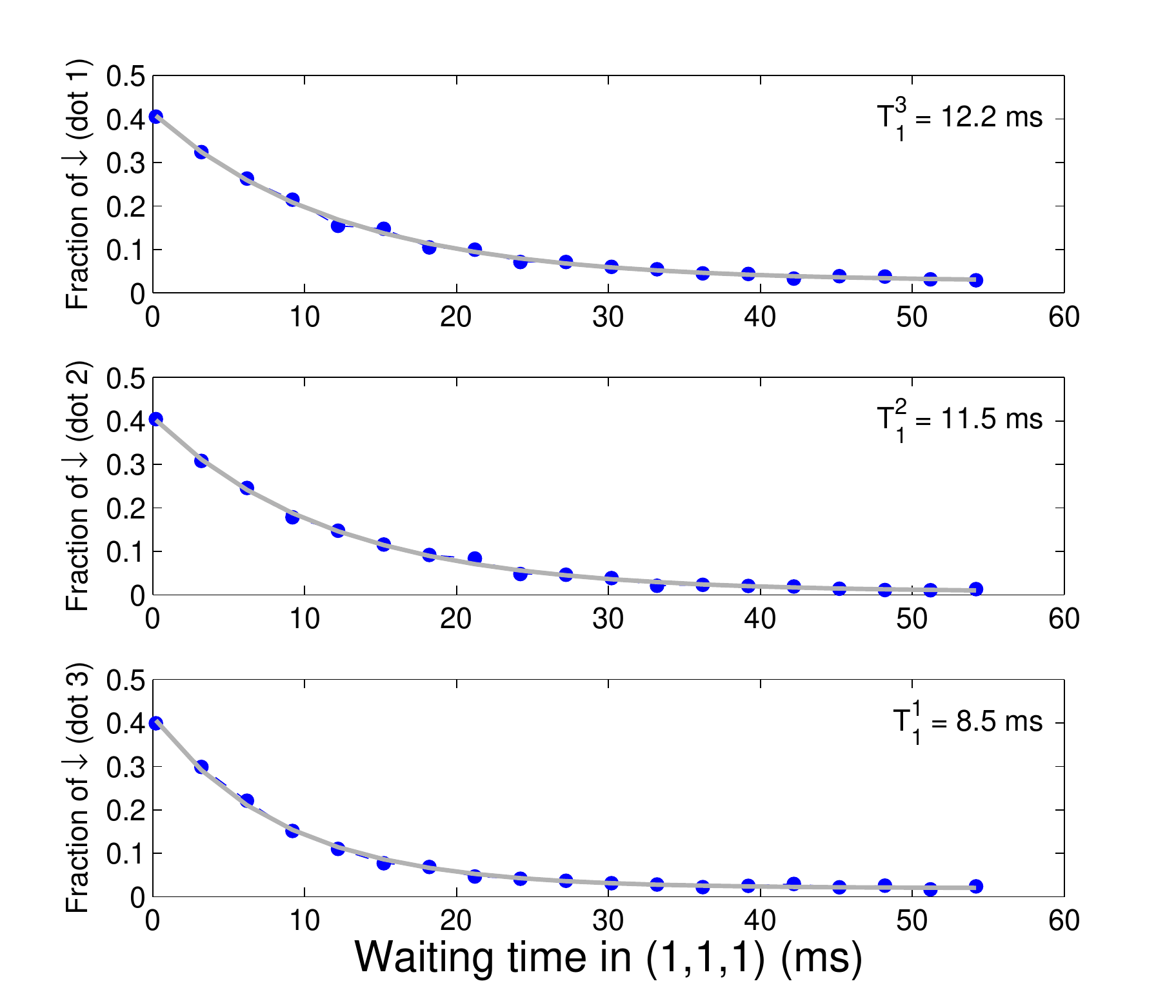}
		\caption{Single dot relaxation as a function of waiting time in (1,1,1). Each datapoint is an average of 2000 measurements. Grey lines are a fit to $p^{j} \cdot e^{-t/T_{1}^{j}} + \alpha^{j} $.}
		\label{fig:figS_T1}
	\end{figure*}

	\textbf{Note on the choice of $T_{1}^{j}$}: as we are transferring some of the spin states throughout the array, they will also experience different $T_{1}$'s depending on their location. In the current scheme each spin is waiting most of the time in its starting dot, so for the calculation of $\eta^{j}$ we take $T_{1}^{j}$. During the read-out itself, so for the calculation of $\Xi^{j}$ and $\Lambda^{j}$ we always use $T_{1}^{j}=T_{1}^{1}$. Fig.~S\ref{fig:figS_T1} shows the three measured $T_{1}^{j}$.
	
	\begin{figure*}[h!]
		\centering
		\includegraphics[width=1.0\textwidth]{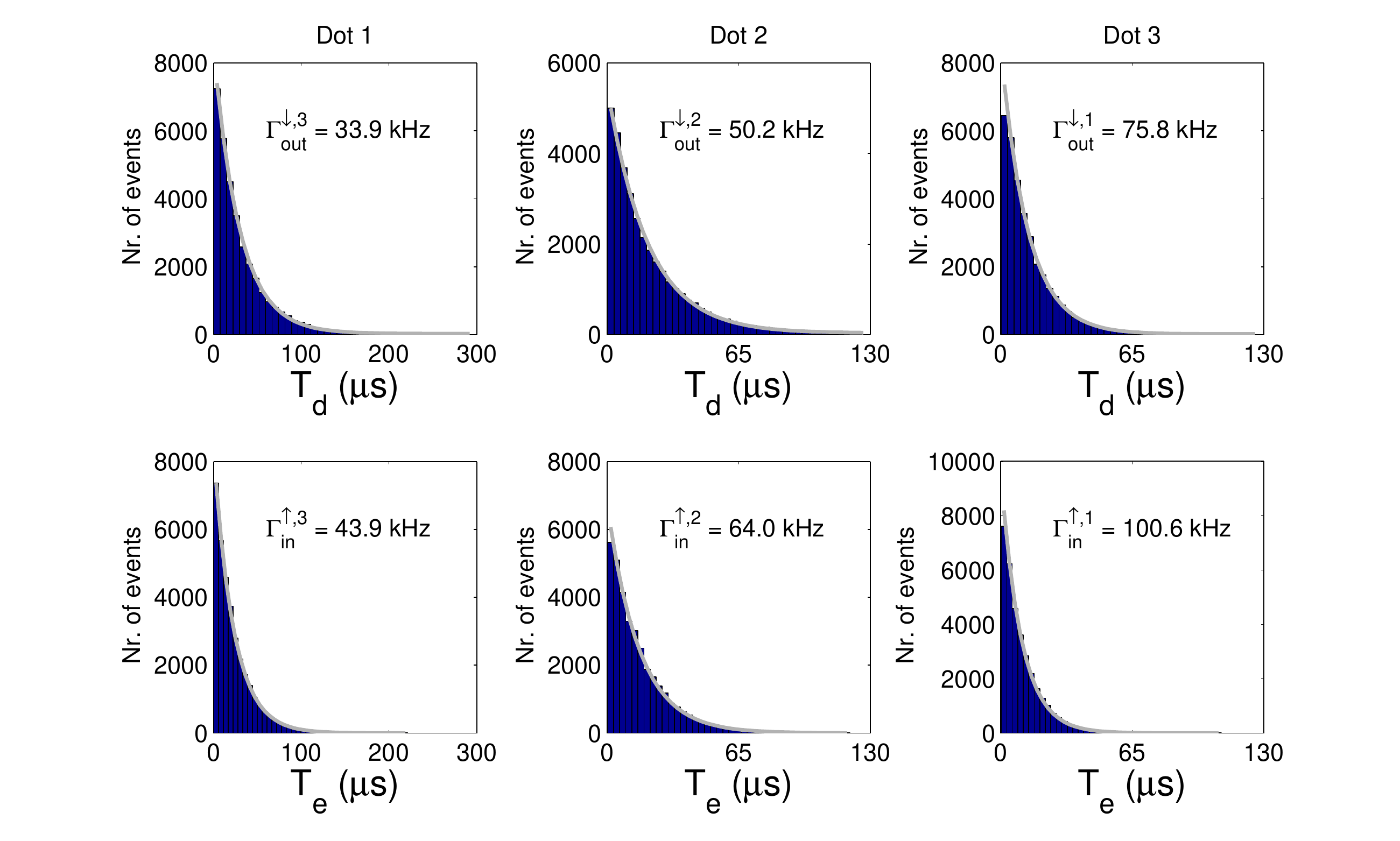}
		\caption{Top row: histograms showing the distribution of the time $T_{d}$ it takes a spin-down electron to tunnel out at the read-out position of each respective dot for the measurement shown in Fig.~S\ref{fig:figS_T1}. The line is an exponential fit from which we determine the decay rate given by $\Gamma_{out}^{\downarrow,j}+\frac{1}{T_{1}^{j}}$. Bottom row: histograms showing the distribution of the time $T_{e}$ it takes a spin-up electron to tunnel back into the empty dot. The grey line is an exponential fit from which we can extract the decay rate given by $\Gamma_{in}^{\uparrow,j}$.}
		\label{fig:figS_tunnel_rates}
	\end{figure*}

	\subsection{Determination of the tunnel rates}
	For calculating the spin-down read-out fidelities, various tunneling rates need to be known. We can obtain $\Gamma_{out}^{\downarrow,j}$ and $\Gamma_{in}^{\uparrow,j}$ from the measurements of the histograms of the time it takes a spin-down electron to tunnel out ($T_{d}^{j}$) and of the time it takes a spin-up electron to tunnel back into the quantum dot ($T_{e}^{j}$), see Fig.~S\ref{fig:figS_tunnel_rates}. We estimate $\Gamma_{out}^{\uparrow,j}$ from the measured value of $\alpha^{j}$, which characterizes the spin up fidelities and can be expressed as
	\begin{equation}
	\alpha^{j}  = \int\limits_{0}^{T_{R}^{j}-0.7 \mu s} \Gamma_{out}^{\uparrow,j} e^{-\Gamma_{out}^{\uparrow,j}t}  dt \int\limits_{0}^{\infty} \Gamma_{in}^{\uparrow,j} e^{-\Gamma_{in}^{\uparrow,j} \tau} B^{j}(\tau) d\tau +\epsilon^{j}
	\end{equation}  
	
	Here $\epsilon^{j}$ is the signal processing error which is found to $\leq 0.1$ \% (to be discussed in the section on the characterization of the measurement bandwidth). 
	
	\subsection{Characterization of the measurement bandwidth}
\begin{figure*}[h!]
		\centering
		\includegraphics[width=0.5\textwidth]{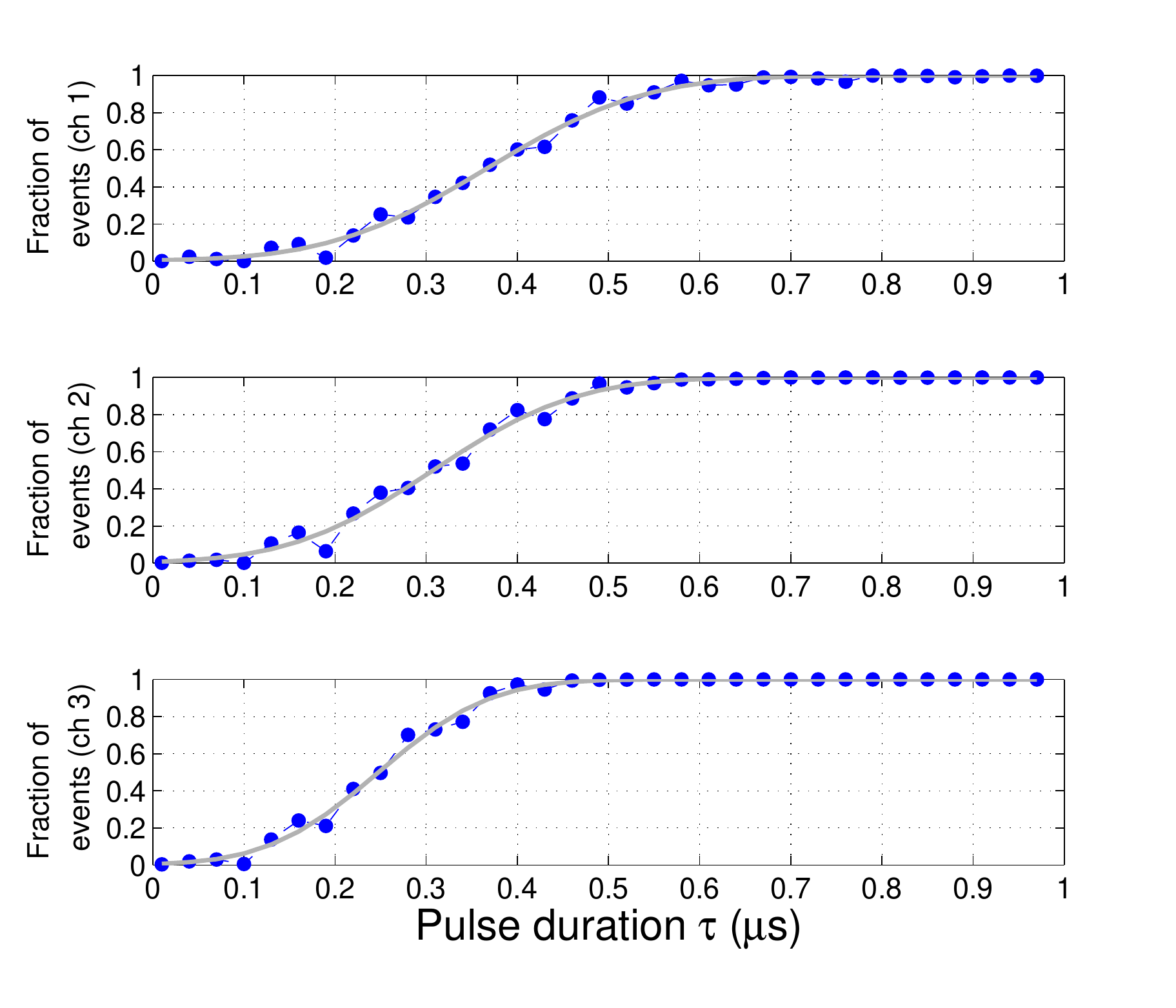}
		\caption{The probability of detecting a pulse with duration $\tau$ for each of the three read-out channels ($B^{j}(\tau)$). Every read-out stage is monitored by a separate input channel of the FPGA. The output of the demodulation box is low-pass filtered at 1 MHz. The sensitivity of the SD to the three different read-out stages is slightly different as they occur at different detunings of the plunger gates. Each datapoint is an average of 1000 measurements.}
		\label{fig:figS2}
	\end{figure*}

	The signature of a spin-down electron tunneling out of the dot followed by a spin-up electron tunneling back into the dot is a pulse in the SD signal. Due to the finite measurement bandwidth, we can only detect pulses of sufficient duration and we therefore miss a fraction of the actual tunnel events. We characterize the measurement bandwidth by simulating such electron tunneling events by applying a rectangular pulse to $P_{3}$ with duration $\tau$ and amplitude $A$. The triple dot is tuned deep into Coulomb blockade such that the applied pulse cannot result in a charge transition of any of the quantum dots. Due to the capacitive coupling of $P_{3}$ to the SD, a rectangular pulse will be created in the SD signal. The amplitude $A$ is chosen such that the height of the pulse in the SD signal equals the signal from electron tunneling events for each respective read-out stage. The pulse duration is varied from 0.0 to 1.0 $\mu s$. Fig.~S\ref{fig:figS2} shows the probability of detecting an event of duration $\tau$, $B^{j}(\tau)$, as a function of the pulse duration for each of the three read-out channels. Using the formulas derived in \emph{Analytic expression for the fidelity} we calculate the fidelities numerically by using the measured probabilities $B^{j}(\tau)$; the result is shown in Table S~\ref{tab:overview_fidelities}. The fraction of the detected events at $\tau=0$ s gives the signal processing error, $\epsilon^{j}$, which corresponds to the fraction of traces in which an event is detected in the SD signal due to noise in the SD-signal during the read-out stage. 
	
	\begin{table}[h!]
		\centering
		\footnotesize
		\begin{tabular}{ |c | c |c | c | c |c |}
			\hline
			dot nr. & $T_{1}$ (ms) (worst,best) & Spin-down fidelity (\%) (worst,best) & Spin-up fidelity (\%) (worst,best) & Orbital splitting (meV) \\ \hline
			1 & 12.2 (11.4, 13.1) & 94.1 (93.8, 94.5)  	& 97.4 (96.7, 98.0) & 2.0 \\ \hline
			2 & 11.5 (10.9, 12.2)& 95.8 (95.6, 96.1) 	& 99.3 (98.8, 99.9) & 1.8 \\ \hline
			3 & 8.5  (8.0, 9.0)& 97.4 (97.3, 97.5)  	& 98.0 (97.6, 98.4) & 1.0 \\ \hline
		\end{tabular}
		\caption{Read-out fidelities per dot. Values in brackets show the error margins based on the fits. For completeness the orbital splitting in each of the three dots was also measured using pulsed spectroscopy~\cite{Elzerman2004b}.}
		\label{tab:overview_fidelities}
	\end{table}
	
	\newpage
	
	\section{Detailed information of the applied pulse sequence}
\begin{figure}[!h]
		\centering
		\begin{tabular}{l l}
			\textbf{a} & \textbf{b} \\
			\includegraphics[width=0.5\textwidth]{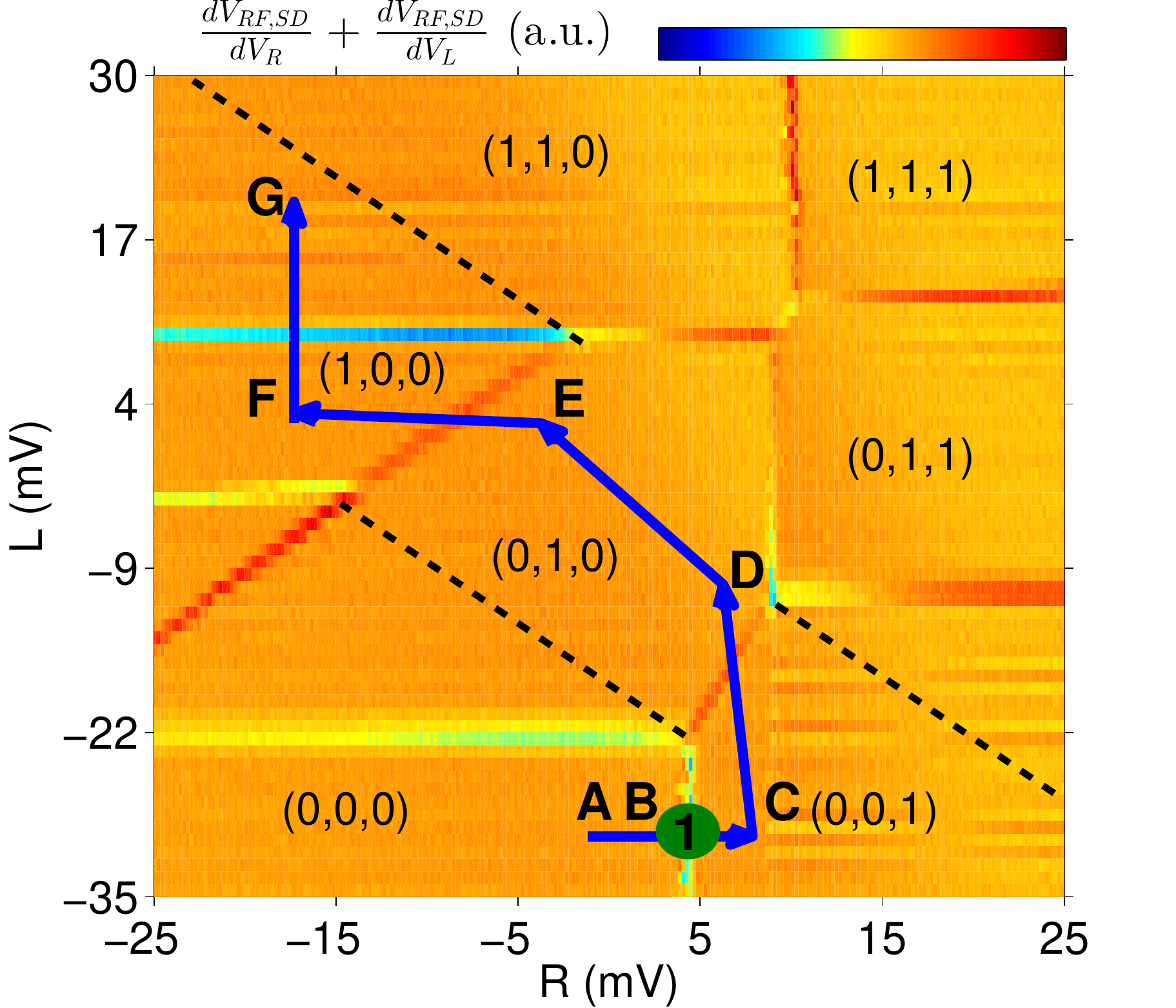} & \includegraphics[width=0.5\textwidth]{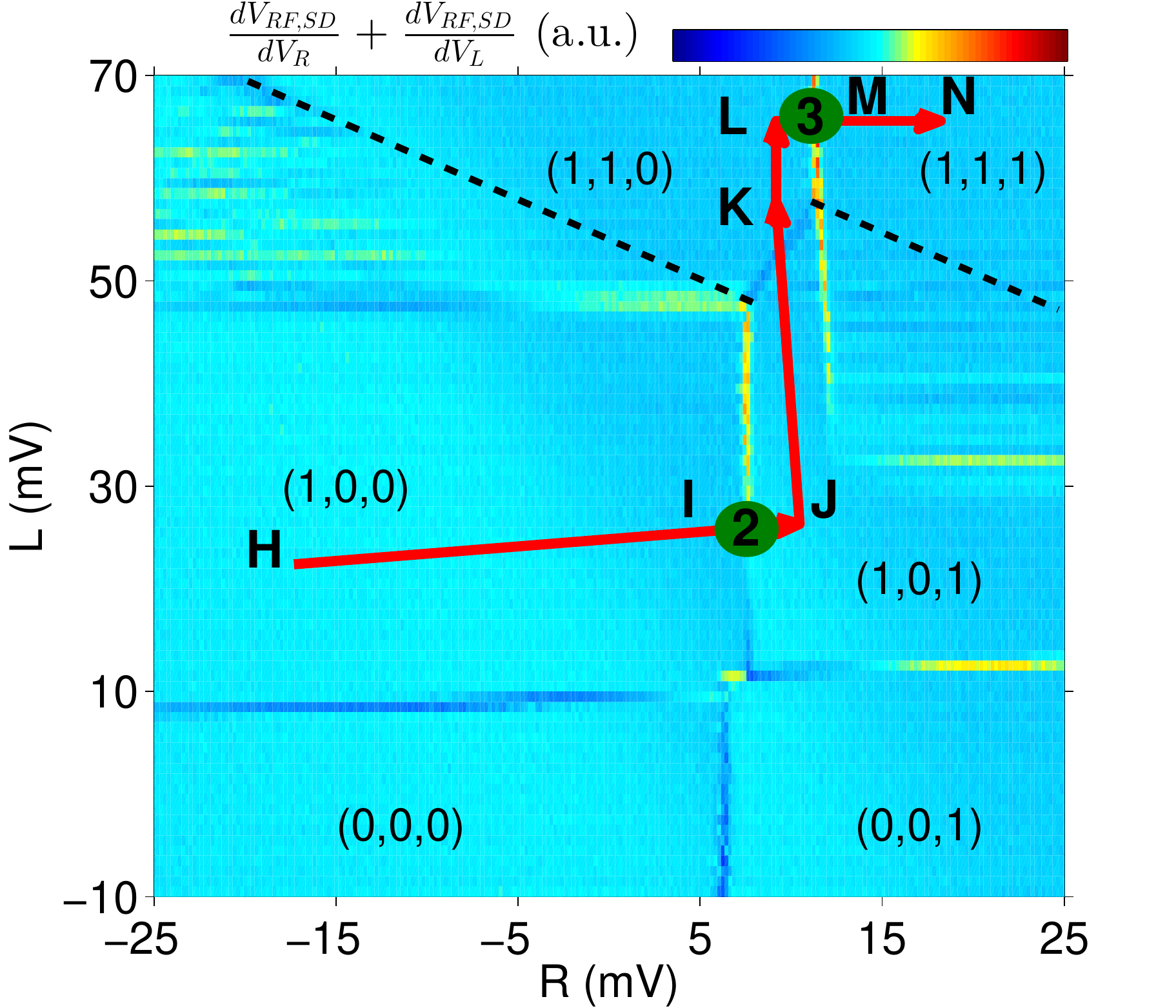} \\
		\end{tabular}
		
		\caption{Charge stability diagram for \textbf{a} $M$ = -42 and \textbf{b} $M$ = -56 from Fig.~1 from the main text with added details for the pulse sequence.}
		\label{fig:honeycombs_detailed_pulse_sequence} 
	\end{figure}

	In this section we give detailed information on the applied pulse sequences. Fig.~S\ref{fig:honeycombs_detailed_pulse_sequence} shows the same charge stability diagram as Fig.~1bc from the main text with additional labeling. Table S~\ref{tab:details_pulse_sequence} gives an explanation including the details of the relevant stages. To correct for slow variations in the dot levels as a function of time (hours timescale), we always calibrate the three read-out stages before each longer measurement such as a complete $T_{1}$ decay ($\sim$ 20.000 datapoints taken after one calibration run, which takes about half an hour). 
	
	\begin{table}[h!]
		\centering
		\small
		\begin{tabular}{ |l | l |}
			\hline
			\textbf{Stage(s)} & \textbf{Purpose and details} \\ \hline
			A & Emptying stage (lasting 1 ms) during which all dots are emptied. \\ \hline
			B $\rightarrow$ C & Loading an electron from the right reservoir into the right dot. We ramp from B to C in 600 $\mu$s\\ 
			&  to load a $\uparrow$-electron, or pulse from B to C to load an electron with random spin. \\ \hline
			C $\rightarrow$ D &  Shuttle electron from the right to the middle dot. Waiting time at D is 10 $\mu$s. \\ \hline
			D $\rightarrow$ E &  Preparing to shuttle to the next dot. Waiting time at E is 1 $\mu$s. \\ \hline	
			E $\rightarrow$ F &  Shuttle electron from the middle to the left dot. Waiting time at F is 10 $\mu$s. \\ \hline
			F $\rightarrow$ G &  Preparing to pulse into the other $M$-plane. Waiting time at G is 1 $\mu$s. \\ \hline
			G $\rightarrow$ H &  Pulse into the other $M$-plane. Waiting time at H is 10 $\mu$s. \\ \hline
			H $\rightarrow$ I &  Preparing to load a new electron into the right dot. Waiting time at I is 1 $\mu$s.\\ \hline
			I $\rightarrow$ J & Loading an electron from the right reservoir into the right dot. We ramp from I to J in 400 $\mu$s \\ 
			&  to load a $\uparrow$-electron, or pulse from I to J to load an electron with random spin. \\ \hline
			J $\rightarrow$ K &  Shuttle electron from the right to the middle dot. Waiting time at K is 10 $\mu$s.\\ \hline
			K $\rightarrow$ L &  Preparing to load a new electron into the right dot. Waiting time at L is 1 $\mu$s.\\ \hline
			L $\rightarrow$ M & Loading an electron from the right reservoir into the right dot. We ramp from L to M in 400 $\mu$s \\ 
			& to load a $\uparrow$-electron, or pulse from L to M to load an electron with random spin. \\ \hline
			M $\rightarrow$ N & Optional stage to perform EDSR deeper into (1,1,1) to prevent photon-assisted tunneling with the\\
			&  reservoirs during the applied FM-burst. \\ \hline
			Read-out stage 3 &  Gate voltage pulses larger than $\sim$2 mV produce a spike in the RF-read-out signal. To prevent false \\
			& events, we therefore first pulse to a position close to the read-out stage (2 mV more positive in $R$)\\ & and wait for 2 $\mu$s. Only then, the RF-signal is unblanked and next we pulse into the read-out \\ & configuration for 130 $\mu$s. \\ \hline		
			L & Empty the right dot for 70 $\mu$s. \\ \hline
			L $\rightarrow$ K $\rightarrow$ J & Shuttle the center electron to the right dot. Waiting time at J is 10 $\mu$s. \\ \hline
			Read-out stage 2 & Similar as Read-out stage 3. \\ \hline
			I  &  Empty the right dot for 100 $\mu$s. \\ \hline
			I $\rightarrow$ C & Path of I to C through all intermediate points with similar times as for the loading sequence. \\ \hline
			Read-out stage 1 & Similar as Read-out stage 3, except we now stay in the read-out configuration for 300 $\mu$s. \\ \hline
			Compensation stage & At the end of the pulse sequence we add a compensation stage of 10-45 ms that ensures that the \\
			& total DC-component of the pulse is zero. This prevents unwanted offsets of the dot levels due to the\\
			& bias tees. \\ \hline 				
		\end{tabular}
		\caption{Detailed explanation of the applied pulse sequence as described by Fig.~S\ref{fig:honeycombs_detailed_pulse_sequence}. Unless noted otherwise, we always apply pulses from one point to the other. The total duration of this sequence varies between 3.3 and 54 ms, not including the compensation stage at the end.}
		\label{tab:details_pulse_sequence}
	\end{table}
	
	\clearpage
	
	\section{Estimation of the error rate during shuttling}
	In the main text we give estimates for the error rate during shuttling caused by (1) charge exchange with the reservoir and (2) random spin flips caused by the hyperfine interaction. In this subsection we will derive the models used to quantify these errors.
	
	\subsection{(1) Charge exchange with the reservoir}
\begin{figure*}[h!]
		\centering
		\includegraphics[width=0.4\textwidth]{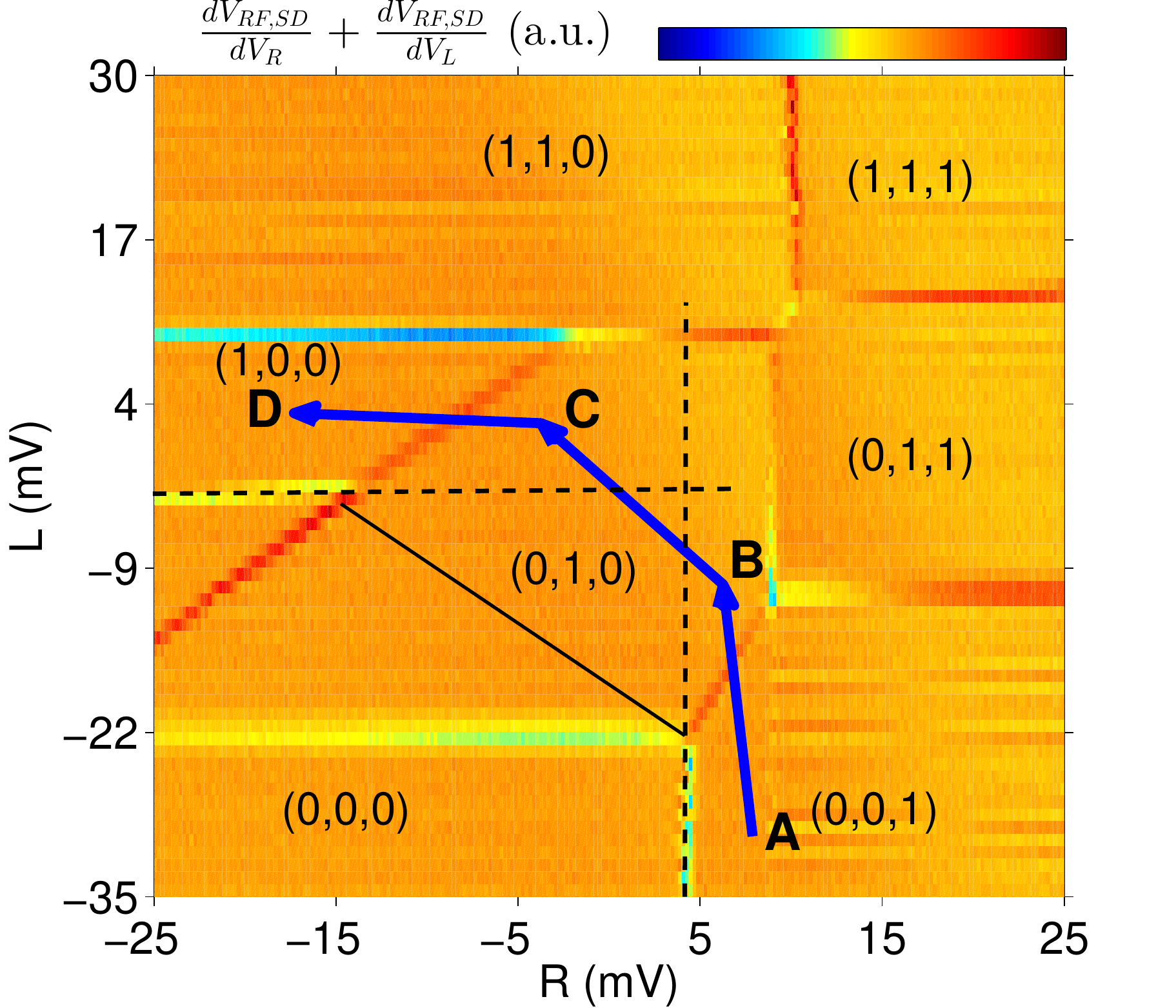}
		\caption{Pulse scheme for the multiple shuttling experiment. We first load a random spin at position A from the right reservoir. Next we go back and forth to A by applying the following pulse sequence: A$\rightarrow$B$\rightarrow$C$\rightarrow$D$\rightarrow$C$\rightarrow$B$\rightarrow$A. Each run  of going back and forth corresponds to four interdot tunnel events. The black dashed lines are extensions of the dot-reservoir charge transition lines. }
		\label{fig:figSmultiple_shuttling_scheme}
	\end{figure*}

	Fig.~S\ref{fig:figSmultiple_shuttling_scheme} shows the pulse scheme of the multiple shuttling experiment in more detail. The probability that pulsing from A to B will be successful (i.e. the electron is fully transferred to B) can be calculated as follows:
	
	\begin{equation}
	P_{A \rightarrow B}(t_{shuttle}) = \int_{0}^{t_{shuttle}}\Gamma_{A \rightarrow B} e^{-\Gamma_{A \rightarrow B}t} dt
	\end{equation}
	
	where $t_{shuttle}$ is the time the electron waits in B before going to C, and $\Gamma_{A \rightarrow B}$ is the interdot tunnel rate between the middle and right dot at that specific detuning. In this experiment the waiting time at each point A, B, C, D is equal to $t_{shuttle} = \frac{t_{CCD}}{1.5 \cdot n_{hops}} $.
	
	In the case that the electron was not successfully transferred to (0,1,0) at B, an error might occur when we continue pulsing to C. Pulsing from B to C crosses the extension of the right dot-reservoir charging line (dashed vertical line in Fig.~S\ref{fig:figSmultiple_shuttling_scheme}). This implies that the following two processes can occur to the electron that is still in the right dot whilst the detuning has already been pulsed to C: (1) the electron will shuttle from the right to the middle dot with rate $\Gamma_{A \rightarrow C}$, (2) the electron will tunnel out to the right reservoir with rate $\Gamma_{reservoir,R}$, and a new electron is eventually loaded, but its spin will not be correlated with that of the initial electron, introducing  an error. The probability that an error occurs going from B to C can therefore be expressed as:
	
	\begin{equation}
	P_{error,B \rightarrow C} = \frac{\Gamma_{reservoir,R}}{\Gamma_{A \rightarrow C} + \Gamma_{reservoir,R}}
	\end{equation}
	Leading to:
	\begin{equation}
	P_{error,A \rightarrow C} = (1-P_{A \rightarrow B}(t_{shuttle})) \cdot P_{error,B \rightarrow C}
	\label{eq:error}
	\end{equation}
	
	In the following, we will neglect the possibility that none of the above two processes occurred at C. We will also neglect the possibility of an unsuccessful tunnel event from  $C \rightarrow D$. This would lead to a second-order correction to the error in the reversed pathway  $D \rightarrow B$. The reversed pathway can then be expressed in a similar way as eq. (\ref{eq:error}). We measured that $\Gamma_{reservoir,R} \approx 20$ kHz, $\Gamma_{reservoir,L} \approx 10$ kHz and $\Gamma_{D \rightarrow B} > \Gamma_{A \rightarrow C} > 1$ MHz (lower bound, limited by the 	measurement bandwidth). The error should therefore be dominated by the path going from A to C. To get an upper bound for the error, we will assume from now on that $P_{error} = P_{error,A \rightarrow C}=P_{error,D \rightarrow B}$.\\
	
	To calculate the measurement outcome at the end of the shuttling sequence, we only have to keep track of the last error in each measurement run. If an electron is exchanged for example twice during a run, we will eventually only measure the spin state of the last electron that entered.\\
	Depending on when the last error occurred, we load a spin-down electron with probability $P_{load,\downarrow}$. Its spin state will decay with $T_{1}=T_{1,weighted}$. Please recall that $P_{error}$ is equal to the error per \textit{two} hops as we neglect errors in the processes $C \rightarrow D $ and $B \rightarrow A $.
	
	\begin{eqnarray}
	P_{\downarrow}(n_{hops},t_{CCD}) = P_{load,\downarrow} \left((1-P_{error})^{n_{hops}/2} e^{-t_{CCD}/T_{1,weighted}}\right) + \nonumber \\ 
	P_{load,\downarrow} \left(\sum_{k=1}^{k=n_{hops}/2} P_{error} (1-P_{error})^{k-1} e^{-(t_{CCD}\cdot \frac{(k-1)}{ n_{hops}/2}  )/(T_{1,weighted})} \right)  
	\label{eq:shuttling_experiment}
	\end{eqnarray}
	
	The first term describes the process of no error occurring. The second term adds up all the possible errors starting from the last error happening in the final stage, $k=1$, till the first stage, $k=n_{hops}/2$.\\
	
\begin{figure}[!h]
		\centering
		\small
		\begin{tabular}{l l}
			\textbf{a} & \textbf{b} \\
			\includegraphics[width=0.4\textwidth]{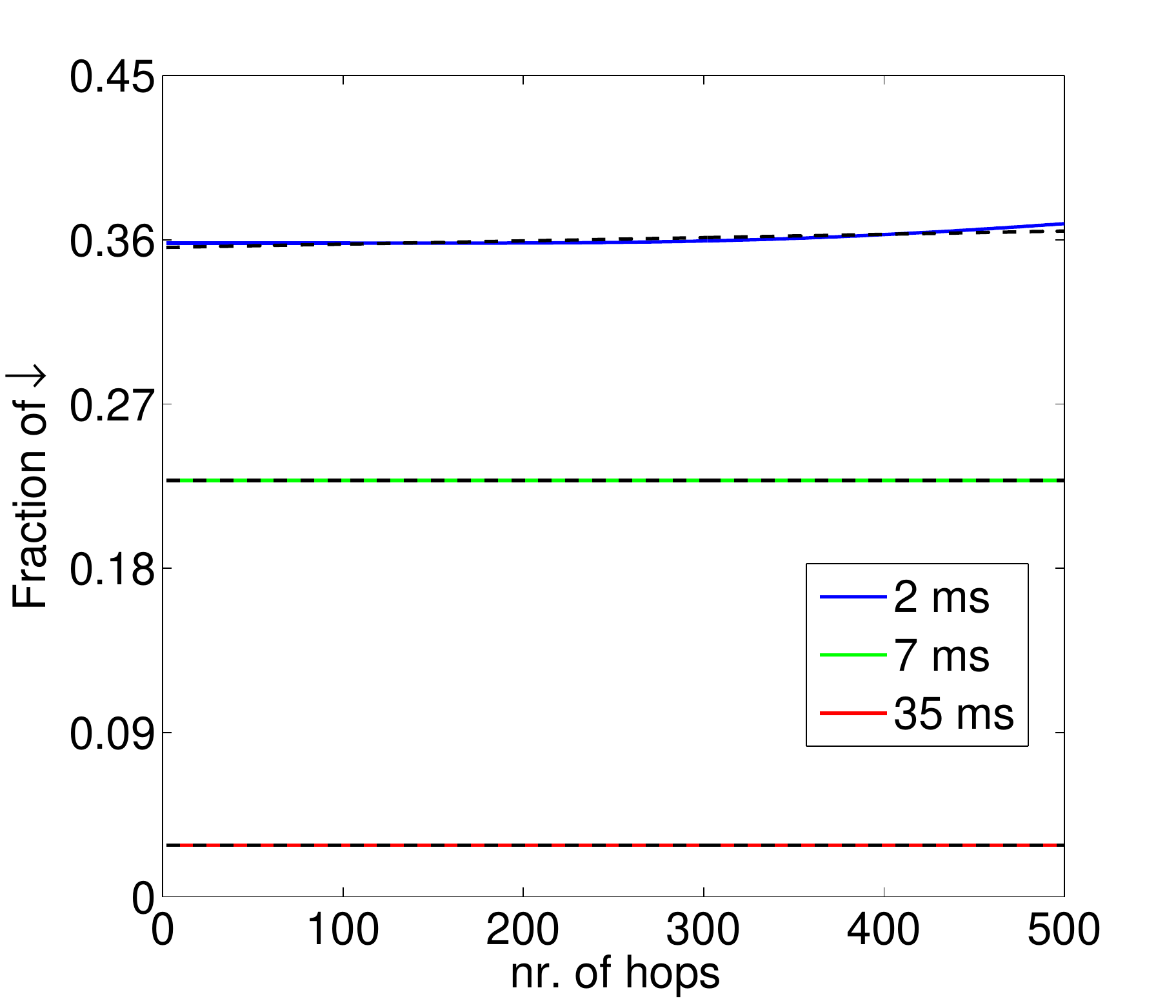} & \includegraphics[width=0.4\textwidth]{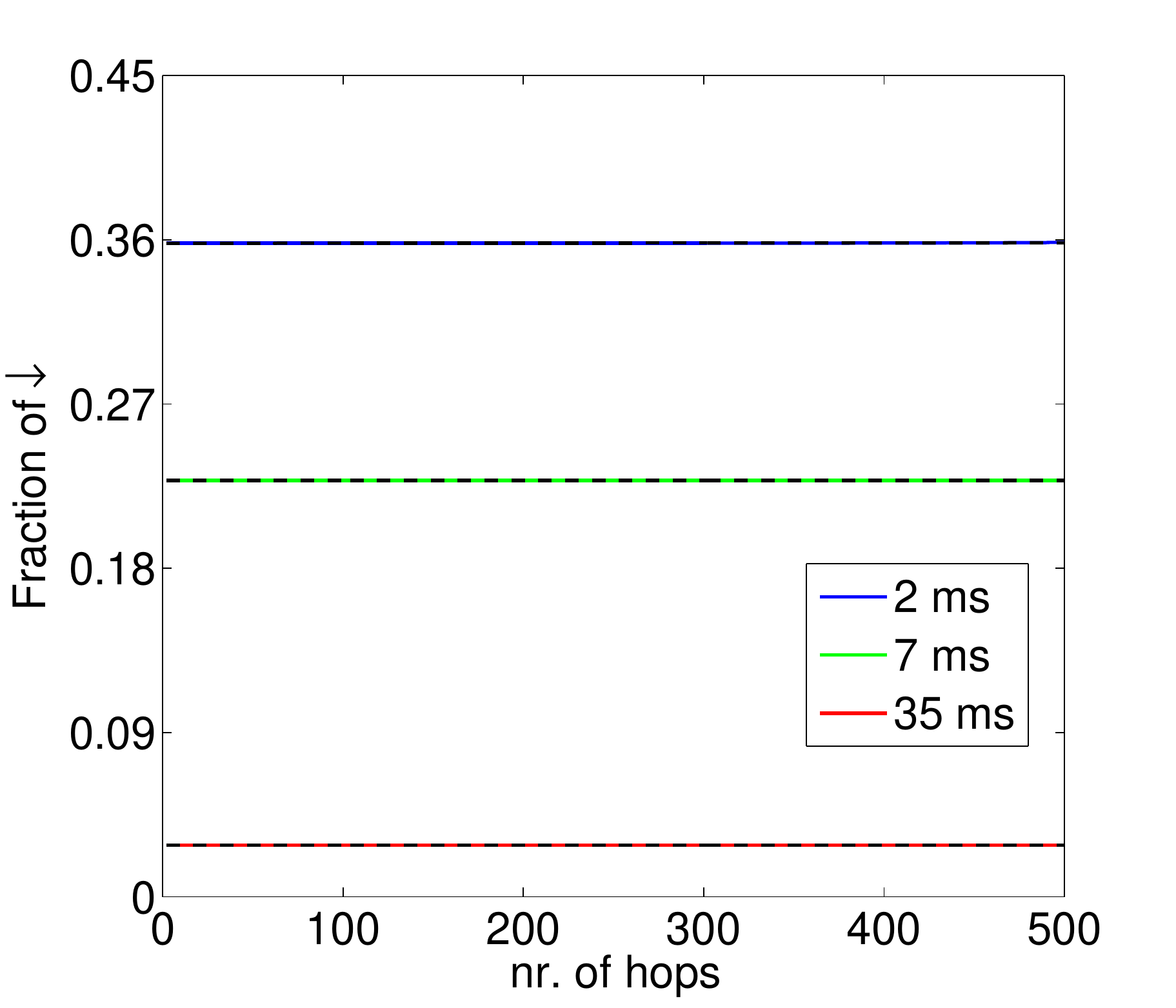}\\
		\end{tabular}
		
		\caption{Outcome of eq. (\ref{eq:shuttling_experiment}) for the same values of $t_{CCD}$ as the main text Fig.~3a for two different values of $\Gamma_{A \rightarrow C}=\Gamma_{D \rightarrow B}$: \textbf{a} 1 MHz and \textbf{b} 2 MHz. In both cases $\Gamma_{reservoir,R} = 20$ kHz. Each curve is linearly fitted (dashed lines) to extract a change in the spin-down fraction per hop.}
		\label{fig:figS_charge_exchange_error}
	\end{figure}

	It was only possible to measure a lower bound for $\Gamma_{A \rightarrow C} > $ 1 MHz. Fig.~S\ref{fig:figS_charge_exchange_error} plots the outcome of eq. (\ref{eq:shuttling_experiment}) for the same values of $t_{CCD}$ as in the main text, Fig.~3a, for two different values of $\Gamma_{A \rightarrow C}=\Gamma_{D \rightarrow B}$: 1 MHz and 2 MHz. Each curve is fitted linearly to extract a change in the spin-down fraction per hop. This linear approximation is only valid for our current regime of small error rates. For large error rates the spin-down fraction will increase non-linearly and eventually form a plateau at the value of $P_{load,\downarrow}$.\\
	The fitted change in the spin-down fraction per hop is $<10^{-8}$ for $t_{CCD} \geq 7$ ms, so we only quote the values for $t_{CCD} = 2$ ms: $18 \cdot 10^{-6}$  ($\Gamma_{A \rightarrow C}=1$ MHz) and  $0.5 \cdot 10^{-6}$ ($\Gamma_{A \rightarrow C}=2$ MHz). The experimentally measured values are given in Table S~\ref{tab:overview_slope_charge_error}.
	
	\begin{table}[h!]
		\centering
		
		\begin{tabular}{ |r | r |}
			\hline
			$t_{CCD}$ (ms) & change in spin-down fraction per hop  \\ \hline
			2 & -1.7 (-25,21) $\cdot 10^{-6}$ \\ \hline 
			7 & 4.6 (-15,24) $\cdot 10^{-6}$ \\ \hline
			35 & 8.3 (-0.97,18) $\cdot 10^{-6}$ \\ \hline
		\end{tabular}
		\caption{The measured change in spin-down fraction per hop measured from the data in Fig.~3a) of the main text. Values in between brackets indicate the 95\% confidence interval of the linear fit.}
		\label{tab:overview_slope_charge_error}
	\end{table}
	
	Based on this simulation we can conclude that charge exchange with the reservoir is a negligible effect for $t_{CCD}=$ 7 and 35 ms and the probability to successfully shuttle from A to B or D to C will approach 1. For the case of $t_{CCD}=$ 2 ms, the slope extracted from the simulation is within the error bars of the experimentally measured value. It is not clear whether charge exchange with the reservoir is already the limiting mechanism here. Finally, we note that when shuttling electrons along longer arrays, charge exchange with the reservoirs will be negligible, if most dots in the array are in fact not coupled to reservoirs.
	
	\newpage
	
	\subsection{(2) Random spin flips caused by the hyperfine interaction}
	To estimate the error rate caused by the hyperfine interaction we describe what happens to a spin that starts in the spin-up state in the first dot. A similar reasoning applies to a spin that starts as spin-down.\\
	The spin-up state in the first dot can be described as a Bloch vector with length 1 and an angle $\theta_{1}=\arctan(B_{N1,\perp}/B_{ext})$ w.r.t to $B_{ext}$ where $B_{N1,\perp}=\sqrt{\left\langle \left(B_{N}^{x,y} \right)^{2} \right\rangle } \approx 5$ mT is the magnitude of the local nuclear field orthogonal to $B_{ext}$ inside dot 1. In the next dot, the spin will experience a different nuclear field: $B_{N2,\perp}$. The Bloch vector will now start rotating along a new axis given by $\theta_{2}=\arctan(B_{N2,\perp}/B_{ext})$ w.r.t to $B_{ext}$. In the current experiment, every step from one dot to the next takes longer than $T_{2}^{*}$. Therefore the Bloch vector will soon dephase into a vector with length $ \cos(\theta_{2}-\theta_{1})$ and point at an angle of $\theta_{2}$ w.r.t $B_{ext}$. For small angles $\Delta \theta = \theta_{2}-\theta_{1} \approx \arctan((B_{N2,\perp}-B_{N1,\perp})/B_{ext}) = \arctan((\sqrt{2}B_{N1,\perp})/B_{ext}) $. Repeating this  $n_{hops}$ times gives a final Bloch vector with length $(\cos(\Delta \theta))^{n_{hops}}$ at an angle $\theta_{\mathrm{final dot}}$. The probability to be in the ground state of the last dot, and thus to be measured as spin-up, is then given by:
	\begin{equation}
	P_{\uparrow} = 0.5 + 0.5 \cdot (\cos(\Delta \theta))^{n_{hops}} \approx 1.0 - \frac{(1-\cos(\Delta \theta))\cdot n_{hops}}{2}
	\label{eq:hyperfine_error}
	\end{equation}
	In the last step of the equation we apply the approximation that $\cos(\Delta \theta) \approx 1$. The error per hop is then estimated by $\frac{(1-\cos(\Delta \theta))}{2} = 1 \cdot 10^{-6}$. 
	
	\section{Improvements to increase the fidelities of the current GaAs experiment}
	This section covers the details of the described improvements for the current GaAs experiment as mentioned in the Discussion of the main text. These improvements combined give the red curve of Fig.~3c of the main text.
	
	To decrease the read-out time we can add a pulsed line to the SD. This allows us to always set the SD to the most sensitive point for each separate read-out position, which should easily improve the measurement bandwidth by a factor of two and thereby reduce the required read-out time by a factor of two. To decrease the emptying-time we propose to add a pulsed line to $RS$, so the tunnel rate between right dot and reservoir can be switched between fast (for emptying) and moderate (for read-out). Combined with larger interdot tunnel couplings both the emptying and the shuttling can take place on the ns timescale. Much faster shuttling would violate the adiabaticity condition, i.e. excitations to higher orbital states would occur~\cite{Taylor2005}. The final improvement involves lowering the magnetic field to 3.0 T (which would require to slightly reduce the electron temperature in order to maintain high read-out fidelities), which already doubles the $T_{1}$ time~\cite{Hanson2007}.

	\clearpage
	
	\section{Virtual gates $L$, $M$ and $R$ and the usage of `fast honeycombs'}
	\textbf{`Virtual gates'}\\
	To control the electron number inside each dot we change the voltages on gates $P_{1}$, $P_{2}$ and $P_{3}$. In practice each gate couples capacitively to all three dots. Changing for example $P_{1}$ which couples mostly to dot 1, will therefore also influence dot 2 and 3. To make selective control of each dot easier it is convenient to correct for this cross-capacitive coupling. This can be done by measuring the cross-capacitance matrix for the three gates recording the influence of each gate on every dot. Inverting this matrix allows you to create honeycomb diagrams with vertical and horizontal charge transitions in a so-called `virtual gate'-space. In such a `virtual gate'-space, the real gates $P_{1}$, $P_{2}$ and $P_{3}$ have been replaced by linear combinations of the three gates which allow the user to change the chemical potential in one dot, without changing it in the neighboring dot.\\
	
	For this experiment the exact relation between  $P_{1}$, $P_{2}$ and $P_{3}$ and the `virtual gates' $L$, $M$ and $R$ used is given by:
	
	\[ \left(\begin{array}{c} L \\ M \\R \\ \end{array}  \right) = \left( \begin{array}{ccc}
	1.56 & 0.0  & 0.40  \\
	0.0  & 1.56 & 0.0 \\
	0.19 & 0.51 & 1.05 \end{array} \right) \left(\begin{array}{c} P_{1} \\ P_{2} \\ P_{3} \\ \end{array}  \right)    \] 
	
	This set of virtual gates did not perfectly correct for cross-capacitances, as the capacitances turn out to slightly vary themselves with the gate voltages. Working with virtual gates does however give one large freedom to take slices at any angle through the 3D honeycomb diagram that satisfy the requirements for the experiment.\\
	
\begin{figure}[!h]
		\centering
		\begin{tabular}{l l}
			\textbf{a} & \textbf{b} \\
			\includegraphics[width=0.5\textwidth]{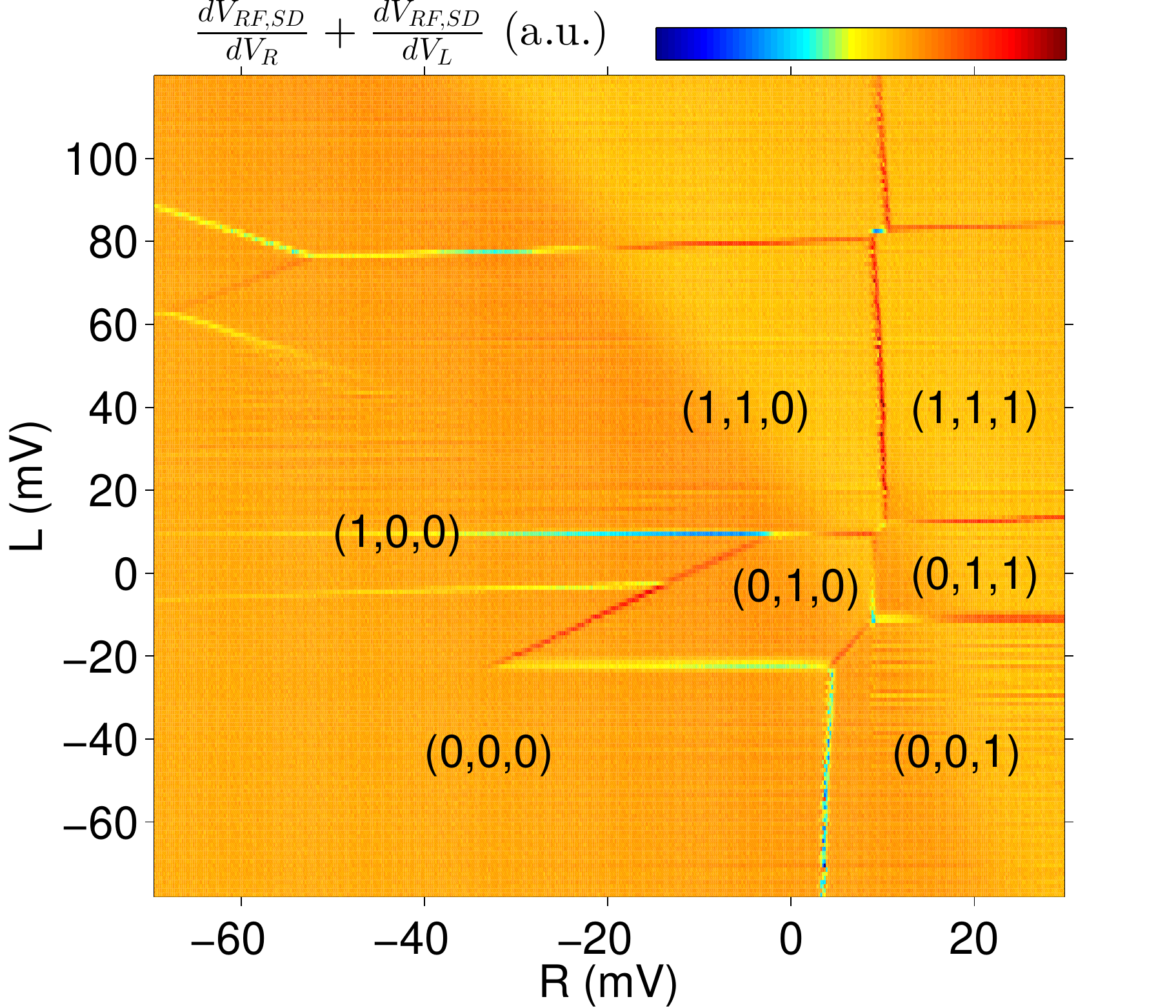} & \includegraphics[width=0.5\textwidth]{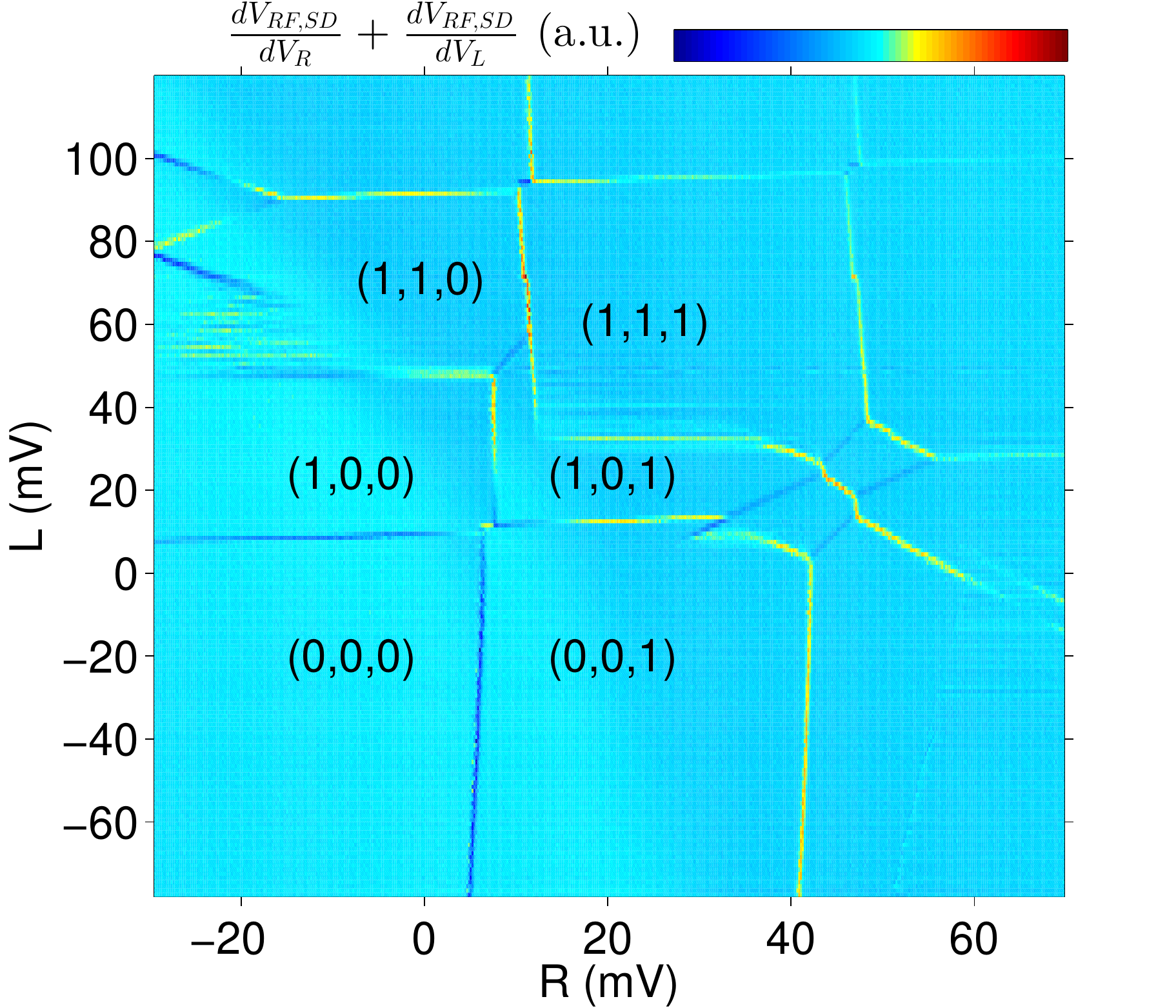} \\
		\end{tabular}
		
		\caption{Charge stability diagram for \textbf{a} $M$ = -42 and \textbf{b} $M$ = -56 shown over a larger range than in Fig.~1.}
		\label{fig:large_honeycombs} 
	\end{figure}

	\textbf{`Fast honeycombs'}\\
	All charge stability diagrams shown in this work have been taken in so-called `fast honeycomb' mode. Using the bias-tees connected to $P_{1}$, $P_{2}$ and $P_{3}$ it is possible to step one of them `slowly' using a DAC and apply a triangular ramp on the other using the AWG. This significantly speeds up the measurements compared to stepping both gates using DAC's. We have always plotted the reverse sweep, i.e. the voltage on horizontal axis is swept from positive to negative (with a rate of 100 mV/44 ms). The fading of the middle dot charging line can be explained through the small interdot tunnel couplings. Fig.~S\ref{fig:large_honeycombs} depicts the same stability diagram as in Fig.~1 of the main text over a larger scan range.   
	
	\clearpage
	
	\section{EDSR spectra of each dot and spin-down initialization efficiencies}
\begin{figure*}[!h]
		\centering
		\includegraphics[width=0.5\textwidth]{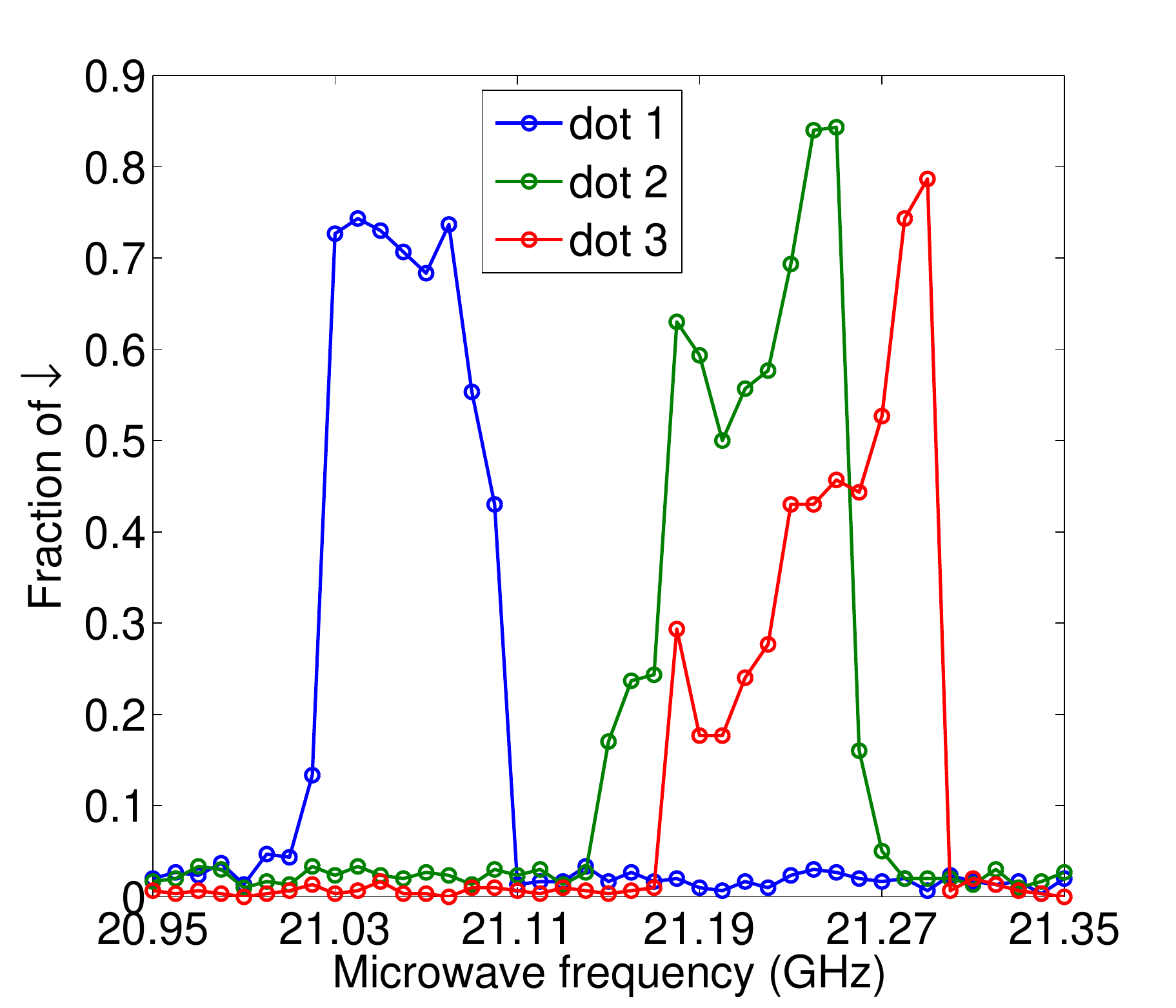}
		\caption{EDSR spectra for each dot. We load $\uparrow \uparrow \uparrow$ inside (1,1,1) and apply a frequency-modulated(FM)-burst to adiabatically invert the spin(s). Afterwards we read out the spin state. Each datapoint is an average of 300 measurements. FM-depth = 60 MHz, burst time = 500 $\mathrm{\mu}$s. Power applied to the gate is -29 dBm (excluding corrections for coax attenuation). The frequency is stepped from 21.35 GHz to 20.95 GHz}
		\label{fig:figS_EDSR_spectra}
	\end{figure*}

	Fig.~S\ref{fig:figS_EDSR_spectra} shows EDSR spectra for each dot measured in a similar way as in~\cite{Shafiei_thesis}.  Despite similar $g_{1}$-factors, time-variations of the local nuclear field still give rise to stable configurations where we can selectively address dot 2 and 3 in the (1,1,1)-regime. To perform such a measurement, the read-out sequence is started and then we manually tune the EDSR-frequency until we find a value that only addresses dot 3. We managed to find configurations stable for more than 30 minutes. We suspect there is some interplay between the electron and nuclear spins that leads to these stable points~\cite{ Vink2009}. This requires further study that is outside the scope of this paper. For completeness Table S~\ref{tab:overview_g_factors} gives an overview of the measured $g$-factors in each dot.\\
	
	Although we use adiabatic passage to create the spin-down states, the initialization efficiency is still limited on average to 76\%. We attribute this non-perfect efficiency to two causes: (1) to achieve selective addressing of dot 2 and 3 (which are closest in $g$-factor) we on purpose slightly detune from the ideal spin-orbit mediated EDSR (SO-EDSR) resonance condition to prevent cross-talk. (2) At 3.51 T, the three hyperfine-mediated-EDSR resonance conditions are detuned by only 26 MHz to 46 MHz from the SO-EDSR condition~\cite{Shafiei2013}. Each resonance frequency by itself is capable of (partially) inverting the spin during a frequency chirp. The used FM-depth of 60 MHz could thus sometimes span several resonances that partially cancel each other.
	
	\begin{table}[h!]
		\centering
		\small
		\begin{tabular}{ |c | c |c |}
			\hline
			dot nr. & $g_{1}$ & $g_{3}$  \\ \hline
			1 & -0.4304 (0.4292, 0.4316) 	& -0.0001853 (-0.0002596, -0.0001109)  \\ \hline 
			2 & -0.4337 (0.4307, 0.4368) 	& -0.0001695 (-0.0003595, 0.0000205)   \\ \hline
			3 & -0.4344 (0.4328, 0.4361)	& -0.0001762 (-0.0002814, -0.0000711)  \\ \hline
		\end{tabular}
		\caption{Measured $g$-factor in each dot based on fits of the form $f_{res}=\frac{g_{1}^{i} \mu_{B} B}{h}+\frac{g_{3}^{i} \mu_{B} B^{3}}{h}$ to the data of Fig.~2i) of the main text. Values in brackets indicate the 95\% confidence interval.}
		\label{tab:overview_g_factors}
	\end{table}

	\clearpage
	
	\section{$T_{1}$ measurements performed for different charge states in the honeycombs}
\begin{figure}[h!]
		\centering
		\begin{tabular}{l l}
			\textbf{a} & \textbf{b} \\
			\includegraphics[width=0.5\textwidth]{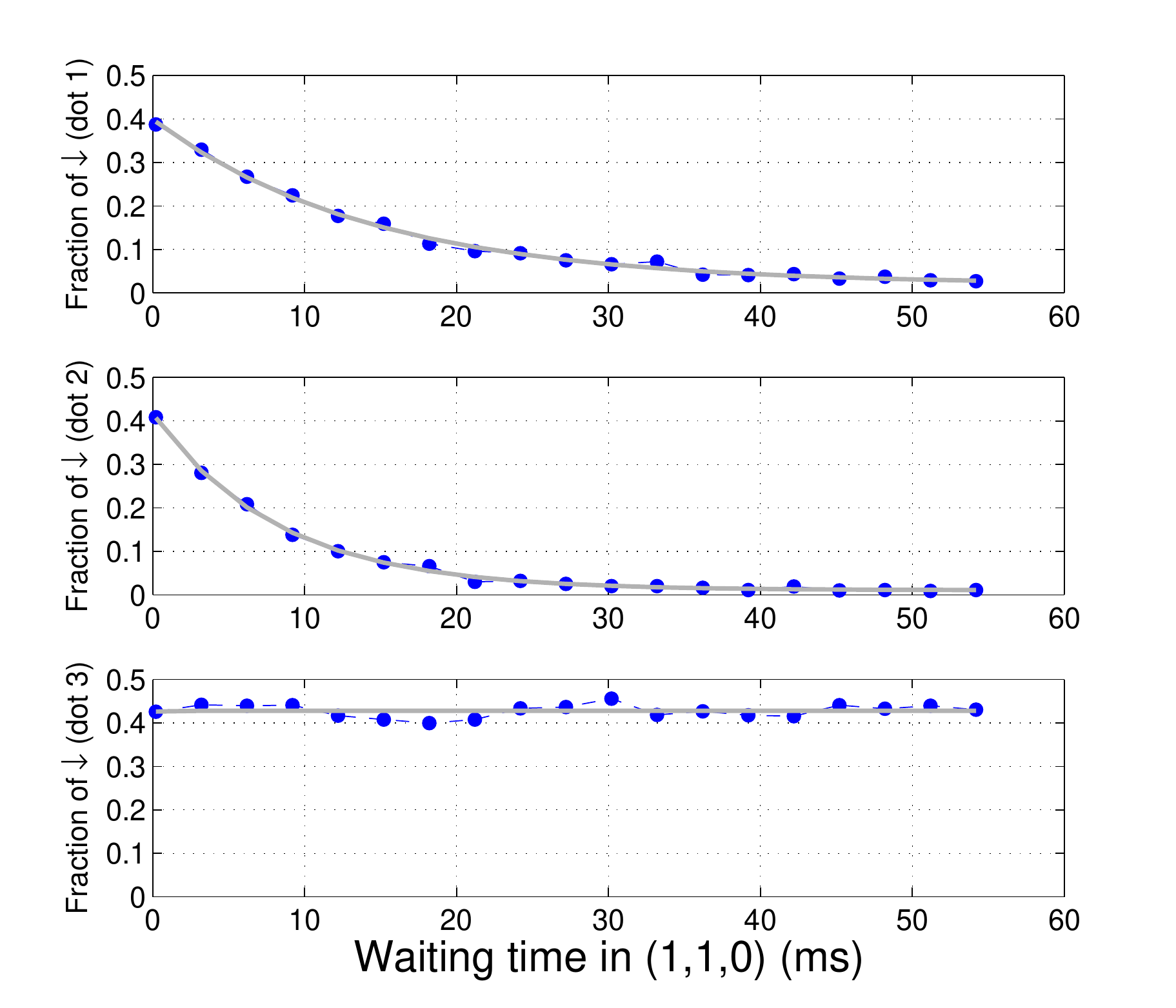} &
			\includegraphics[width=0.5\textwidth]{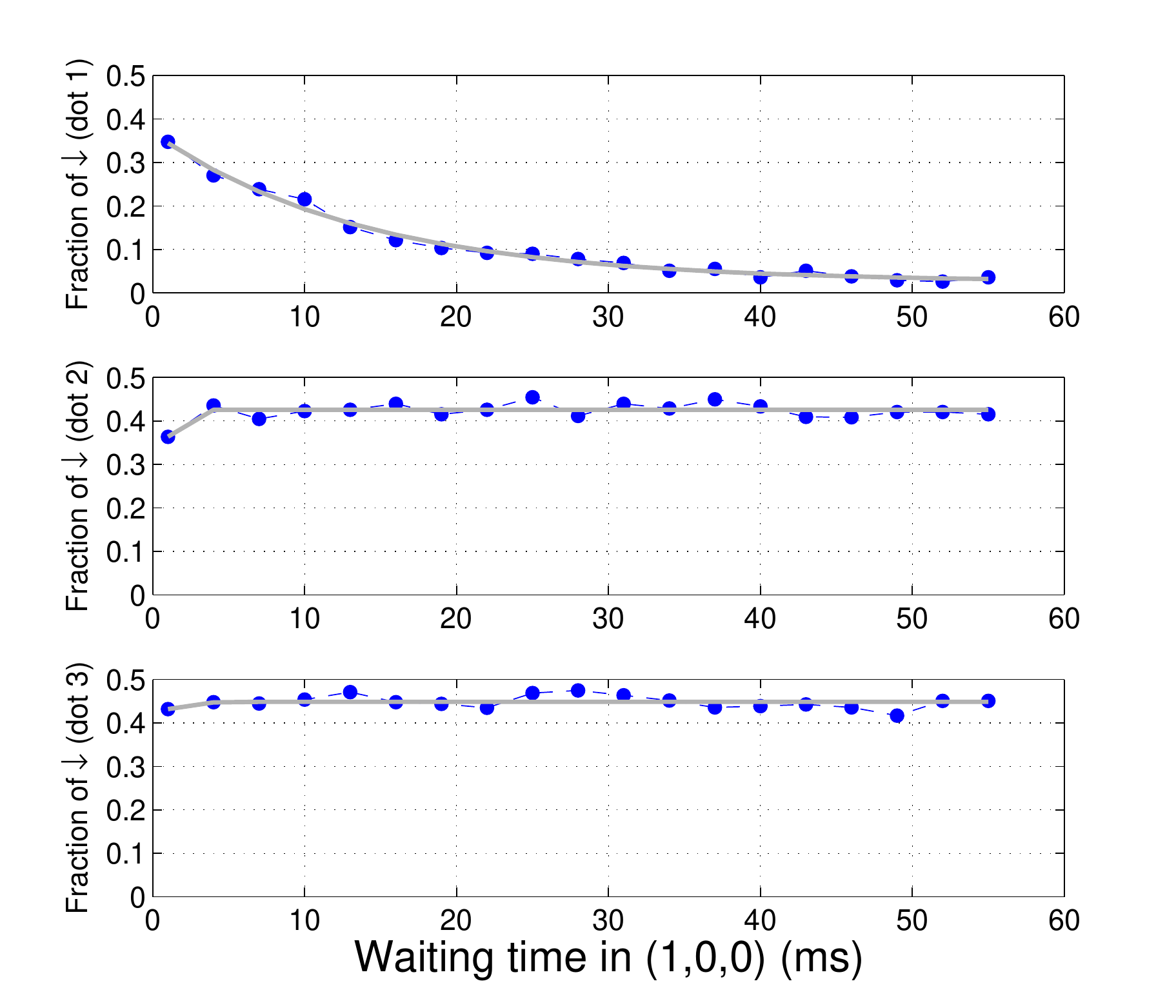}
		\end{tabular}
		\caption{\textbf{a} $T_{1}$ decay in (1,0,1). Fits to the data give $T_{1}^{3}$ = 14.3 ms, $T_{1}^{2}$ = 8.2 ms. The center spin now spends most of its time in dot 3, and therefore also relaxes with the same rate as $T_{1}^{1}$ as measured in (1,1,1). \textbf{b} $T_{1}$ decay in (1,0,0). Fit to the data gives $T_{1}^{3}$ = 13.9 ms.}
		\label{fig:T1_101}
	\end{figure}

	To further verify the proper operation of the single spin CCD we have also performed $T_{1}$ measurements where we do not vary the waiting time in (1,1,1) as shown in Fig.~2 of the main text, but instead change the waiting time in different charge states. In Fig.~S\ref{fig:T1_101} we measure the $T_{1}$ decay of the spins by first loading three random spins in (1,1,1), and next vary the waiting time in (i) (1,0,1) or (ii) (1,0,0). In case (i) the right spin has already been read out before the waiting time, so its spin state will not change as a function of waiting time. The spin that was originally in the center dot is moved to the right dot before the waiting time, so it will now experience a $T_{1}$ time that is more similar to $T_{1}^{1}$. Case (ii) shows no time-dependence for the center and right spin, and does show the $T_{1}$ decay as expected for the left dot. \\
	
\end{document}